# Reconstructing the Population Genetic History of the Caribbean

Andrés Moreno-Estrada[1], Simon Gravel[1,2], Fouad Zakharia[1], Jacob L. McCauley[3], Jake K. Byrnes[1,4], Christopher R. Gignoux[5], Patricia A. Ortiz-Tello[1], Ricardo J. Martínez[3], Dale J. Hedges[3], Richard W. Morris[3], Celeste Eng[5], Karla Sandoval[1], Suehelay Acevedo-Acevedo[6], Juan Carlos Martínez-Cruzado[6], Paul J. Norman[7], Zulay Layrisse[8], Peter Parham[7], Esteban González Burchard[5], Michael L. Cuccaro[3], Eden R. Martin[3*], Carlos D. Bustamante[1*]

[1]Department of Genetics, Stanford University School of Medicine, Stanford, CA, USA
[2]Department of Human Genetics and Genome Quebec Innovation Centre, McGill University, Montreal, QC, Canada
[3]Center for Genetic Epidemiology and Statistical Genetics, John P. Hussman Institute for Human Genomics, University of Miami Miller School of Medicine, Miami, FL, USA
[4]Ancestry.com DNA, LLC, San Francisco, CA, USA
[5]Department of Bioengineering and Therapeutic Sciences, University of California San Francisco, CA, USA
[6]Department of Biology, University of Puerto Rico at Mayaguez, Puerto Rico
[7]Department of Structural Biology, Stanford University School of Medicine, Stanford, CA, USA
[8]Center of Experimental Medicine "Miguel Layrisse", IVIC, Caracas, Venezuela
*Shared senior authorship and co-corresponding authors

**Running title:** Reconstructing Human Migrations into the Caribbean

**Contributions:** Conceived and designed experiments: ERM, MLC, JLM, AM, CDB. Performed the experiments: JLM, RJM, DJH, CE. Analyzed the data: AM, SG, FZ, JKB, CRG, PO, SA. Contributed reagents/materials/analysis tools: ERM, MLC, JLM, RWM, KS, JCM, PJN, ZL, PP, EGB, CDB. Wrote the paper: AM, SG, FZ, JKB, CRG, CDB.

**Abstract**

The Caribbean basin is home to some of the most complex interactions in recent history among previously diverged human populations. Here, we investigate the population genetic history of this region, by characterizing patterns of genome-wide variation among 330 individuals from three of the Greater Antilles (Cuba, Puerto Rico, Hispaniola), two mainland (Honduras, Colombia), and Native South American (Yukpa, Bari, and Warao) populations. We combine these data with a unique database of genomic variation in over 3,000 individuals from diverse European, African, and Native American populations. We use local ancestry inference and tract length distributions to test different demographic scenarios for the pre- and post-colonial history of the region. We develop a novel ancestry-specific PCA (ASPCA) method to reconstruct the sub-continental origin of Native American, European, and African haplotypes from admixed genomes. We find that the most likely source of the indigenous ancestry in Caribbean islanders is a Native South American component shared among coastal tribes from Venezuela, Central America, and the Yucatan peninsula, suggesting extensive gene flow across the Caribbean in pre-Columbian times. We find evidence of two pulses of African migration. The first pulse—which today is reflected by shorter, older ancestry tracts—consists of a genetic component more similar to coastal West African regions involved in early stages of the trans-Atlantic slave trade. The second pulse—reflected by longer, younger tracts—is more similar to present-day West-Central African populations, supporting historical records of later transatlantic deportation. Surprisingly, we also identify a Latino-specific European component that has significantly diverged from its parental Iberian source populations, presumably as a result of small European founder population size. We demonstrate that the ancestral components in admixed genomes can be traced back to distinct sub-continental source populations with far greater resolution than previously thought, even when limited pre-Columbian Caribbean haplotypes have survived.

**Author Summary**

Latinos are often regarded as a single heterogeneous group, whose complex variation is not fully appreciated in several social, demographic, and biomedical contexts. By making use of genomic data we characterize ancestral components of Caribbean populations on a sub-continental level and unveil fine-scale patterns of population structure distinguishing insular from mainland Caribbean populations as well as from other Hispanic/Latino groups. We provide genetic evidence for an inland South American origin of the Native American component in island populations and for extensive pre-Columbian gene flow across the Caribbean basin. The Caribbean-derived European component shows significant differentiation from parental Iberian populations, presumably as a result of founder effects during the colonization of the New World. Based on demographic models, we reconstruct the complex population history of the Caribbean since the onset of continental admixture. We find that insular populations are best modeled as mixtures absorbing two pulses of African migrants, coinciding with early and maximum activity stages of the transatlantic slave trade. These two pulses appear to have originated in different regions within West Africa, imprinting two distinguishable signatures in present day Afro-Caribbean genomes and shedding light on the genetic impact of the dynamics occurring during the slave trade in the Caribbean.

**Introduction**

Genomic characterization of diverse human populations is critical to enable multi-ethnic genome-wide association and sequencing studies of complex biomedical traits [1]. The increasing availability of genome-wide data from populations worldwide allows for the reconstruction of population history at finer scales, shedding light on evolutionary processes shaping the genetic composition of peoples with complex demographic histories. This is especially relevant in recently admixed populations from the Americas. Native peoples throughout the American continent suffered a dramatic demographic change triggered by the arrival of Europeans and the subsequent African slave trade. Important progress has been made to characterize genome-wide patterns of these three continental-level ancestral components in admixed populations from the continental landmass [2], and other Hispanic/Latino populations [3], including recent genotyping and sequencing studies involving Puerto Rican samples [4,5,6]. However, no genomic survey has focused on multiple populations of Caribbean descent, and critical questions remain regarding their recent demographic history and fine-scale population structure. Several factors distinguish the Antilles and the broader Caribbean basin from the rest of North, Central, and South America, resulting in a unique territory with particular dynamics impacting each of its ancestral components. First, unlike other regions of the Americas, native pre-Columbian populations suffered dramatic population bottlenecks soon after contact, resulting in the disappearance of many genetic lineages. This challenges the reconstruction of the indigenous population genetic history since extant admixed populations have retained a limited proportion of the native genetic lineages [7]. Second, it is widely documented that the initial encounter between Europeans and Native Americans, such as the first voyages of Columbus, took place in the Caribbean before involving mainland populations. However it remains unclear whether the earlier onset of admixture in the Caribbean translates into substantial differences in the European genetic component of present day admixed Caribbean genomes, compared to other Hispanic/Latino populations impacted by later, and probably more numerous, waves of European migrants. Third, the Antilles and surrounding mainland of the Caribbean were the initial destination for much of the African migration into the Americas during the slave trade, resulting in descendant populations retaining higher levels of African ancestry compared to most inland populations across the continent. Yet, details about the impact of the sub-continental origin of African migrants into the composition of Afro-Caribbean genomes remain greatly under-characterized.

Disentangling the origin and interplay between the ancestral components during the process of admixture will enhance our knowledge about the composition of populations living in the Caribbean and diaspora populations of Caribbean descent, informing the design of next-generation medical genomic studies involving these populations. Here, we present SNP array data for 251 admixed individuals from parent-offspring trios of Caribbean descent sampled in South Florida, including Cubans, Haitians, Dominicans, Puerto Ricans, Colombians, and Hondurans, as well as 79 native Venezuelans sampled along the Caribbean coast. We construct a unique database which includes public and DAC-controlled data on genomic variation from over 3,000 individuals including HapMap [8], 1000 Genomes [6], and POPRES [9] populations, and African [10] and Native American [11] SNP data from diverse sub-continental populations employed as reference panels. We apply admixture deconvolution methods and develop a novel ancestry-specific PCA method (ASPCA) to infer the sub-continental origin of haplotypes along the genome, yielding a finer resolution picture of the ancestral components of present day Caribbean and surrounding mainland populations. Additionally, by analyzing the tract length distribution of genomic segments attributable to distinct ancestries we test different demographic models to reconstruct the recent population history of the Greater Antilles and mainland populations since the onset of inter-continental admixture.

**Results**

*Population Structure of the Caribbean*

To characterize population structure across the Antilles and neighboring mainland populations, we combined our genotype data for the six Latino populations with continental population samples from western Africa, Europe, and the Americas, as well as additional admixed Latino populations (see Table S1). To maximize SNP density, we initially restricted our reference panels to representative subsets of populations with available Affymetrix SNP array data (Figure 1A). Using a common set of ~390K SNPs, we applied both principal component analysis (PCA) and an unsupervised clustering algorithm, ADMIXTURE [12], to explore patterns of population structure. Figure 1B shows the distribution in PCA space of each individual, recapitulating clustering patterns previously observed in Hispanic/Latino populations [3]: Mexicans cluster largely between European and Native American components, Colombians and Puerto Ricans show three-way admixture, and Dominicans principally cluster between the African and European components. Ours is the first study to characterize genomic patterns of variation from (1) Hondurans, which we show have a higher proportion of African ancestry than Mexicans, (2) Cubans, which show extreme variation in ancestry proportions ranging from 2% to 78% West African ancestry, and (3) Haitians, which showed the largest average proportion of West African ancestry (84%). Additional projections of PC1 versus higher PCs are shown in Figure S1.

We use the program ADMIXTURE to fit a model of admixture where an individual's genome is composed of sites from up to $K$ ancestral populations. We explored $K=2$ through 15 ancestral populations (Figure S2) to investigate how assumptions regarding $K$ impact the inference of population structure. Assuming a $K=3$ admixture model, population admixture patterns are driven by continental reference samples with no continental subdivision (Figure 1C, top panel). However, higher $K$s show substantial substructure in all three continental components. Log likelihoods for successively increasing levels of $K$ continue to increase substantially as $K$ increases (Figure S3a) which is not unexpected since higher values of $K$ add more parameters to the model (therefore improving the fit). Using cross-validation we find that $K=7$ and $K=8$ have the lowest predicted error (Figure S3b); for this reason we focus on these two models.

The first sub-continental components that emerge are represented by South American population isolates, namely the three Venezuelan tribes of Yukpa, Warao, and Bari. At higher order Ks, we recapitulate the well documented North–South American axis of clinal genetic variation described by us [13] and others [11,14] as Mesoamerican (Maya/Nahua) and Andean (Quechua/Aymara) populations are assigned to different clusters (Figure S2). Interestingly, Mayans are the only group showing substantially higher contributions from the native Venezuelan components (Figure 1C, bottom panel). Above $K=7$, we observe a North–South European differentiation which is consistent with previous analyses [15,16]. Surprisingly, we observe another European-specific component emerge as early as $K=5$ and remain constant through $K=15$. This component accounts for the majority of the Latino's European ancestry and it only appears in Mediterranean populations, including Italy, Greece, Portugal, and Spain at intermediate proportions (Figure S2). Throughout this paper, we refer to this component as the "Latino European" component and it can be seen clearly in Figure 1C ("black" bars represent the Latino European component, "Red" bars represent the "Northern European", and pink the "Mediterranean" or "Southern European" component). At $K=8$, when the clinal gradient of differentiation between Southern and Northern Europeans appears, the Latino European component is seen only in low proportions in individuals from Portugal and Spain, whereas it is the major European component among Latinos (Figure 1C, bottom panel).

Colombians and Hondurans show considerably higher proportions of Native Venezuelan components, consistent with their geographic proximity. Both Mesoamerican and Andean Native American samples contain considerable amounts of European ancestry, largely due to post-Columbian admixture. Interestingly, the European component in Native Americans is assigned to the Latino-specific component in Mesoamericans (Nahua/Maya) and to the Mediterranean-specific European component in Andeans (Aymara/Quechua). The Latino-specific component could be explained as the result of a founder effect driven by early European settlers, hence this pattern would be compatible with an initial introduction of European segments in native Mesoamericans and a later arrival of European chromosomes into the Andean gene pool.

Our data show a strong signature of assortative mating based on genetic ancestry among Caribbean Latinos as suggested by previous studies [17]. In particular, we see a strong correlation between maternal and paternal ancestry proportions (Figure S4). To assess significance, we compared correlation of ancestry assignments among parent pairs to 100,000 permuted male-female pairs for each continental ancestry. All p-values were highly significant ($p < 0.00001$, Table S2). It should be noted that these tests are not independent since the three components of ancestry by definition must sum to one. Further, apparent assortative mating could be due to random mating within structured sub-populations. To control for this, we performed permutations within sampling localities, and found significant correlations among individuals from every single population, except for Haiti. Although Haitians do show the same trend, with only 2 parent pairs is nearly impossible to assess significance (see Table S2).

*Demographic Inference since the onset of admixture*

An overview of our analytic strategy for characterizing admixed genomes is presented in Figure 2. Due to meiotic recombination, the correlation in ancestry among founder chromosomes is broken down over time. As a consequence, the length of tracts assigned to distinct ancestries in admixed genomes is informative of the time and mode of migration [18]. To explore the population genetic history of the Caribbean since European colonization, we considered the length distribution of continuous ancestry tracts in each of the six population samples. First, we estimated local ancestry along the genome using an updated version of PCAdmix [19] which was trained using trio-phased data from the admixed individuals and three continental reference populations. Next, we characterized the length distribution of unbroken African, European or Native American ancestry tracts along each chromosome for each population. Finally, we applied the extended space Markov model implemented in *Tracts* [20] to compare the observed data with predictions from different demographic models considering various migration scenarios.

The simplest model considers a single pulse of migration from each source population, allowing the admixture process to begin with Native American and European chromosomes, followed by the introduction of African chromosomes. In such a scenario each population contributes migrants at a discrete period in time, and the average length of ancestry tracts is expected to decrease with time after admixture, resulting in an exponential decay in the abundance of tracts as a function of tract lengths. Alternative models include a second pulse of either European or African segments migrating into the already admixed gene pool. Allowing for continuous or repeated migration typically results in a concave log-scale distribution, caused by the increase of longer tracts after the second migration event. Table 1 and Figure 3 summarize the results of the best-fitting migration models for each population based on Bayesian Information Criterion (BIC) comparisons, and Figure S5 shows the full results of all models tested. We observed that multiple pulses of admixture offered a better BIC in all cases.

The best-fit model for Colombians and Hondurans involves admixture between Native Americans and Europeans starting 14 generations ago, followed by a second pulse of European ancestry starting 12 and 5 generations ago, respectively. Of note is that between the first and second pulse of migration in Colombians, the proportion of European ancestry increased from 12.5% to 75% in two generations, implying that the European segments in today's Colombians date back to European gene flow happening in a short period of time, thus tracing back their ancestry to a more limited number of founders compared to other Latino populations.

In contrast with mainland population samples, the best-fit model for all four populations from the Caribbean islands involves older time estimates of the initial contact between Native Americans and Europeans. Namely, 17 generations ago for Cubans and 16 generations ago for Puerto Ricans, Dominicans, and Haitians. Historical records state that the first European colonies in the Antilles were set up soon after the initial contact in 1492 [21], that is ~500 years ago or 16.6 generations ago (considering 30 years per generation [22]), in excellent agreement with our time estimates. Another major distinction is that the model involves a second pulse of African ancestry, occurring between 7 to 5 generations ago, with higher migration rates in Haitians and Dominicans, followed by Cubans and Puerto Ricans.

*Sub-continental Ancestry of Admixed Genomes*

The genomes of admixed populations contain information about both continental and sub-continental population processes. To explore within-continent population structure, we performed PCA on genomic segments of specific continental ancestry. Because the masking out of the other ancestries results in large amounts of missing data, we implemented a novel variation of PCA that allows performing the analysis on the remaining sites alone. Throughout this paper, we refer to this approach as ancestry-specific PCA (ASPCA) and the mathematical details are described in Text S1. We applied this methodology for projecting phased genomic segments of inferred Native American, European, and African continental ancestry onto sub-continental reference panels of parental populations (see diagram in Figure 2). Our implementation is analogous to the subspace PCA (ssPCA) approach by Johnson et al. [23], but it can take advantage of phased data, allowing to accommodate parts of the genome that are heterozygous for ancestry. In the presence of recent admixture, chromosomal ancestry breakpoints dramatically reduce the proportion of the genome that is homozygous for a given ancestry. Therefore, relying on genotypes and restricting to loci estimated to have two copies of a certain ancestry could severely compromise the resolution of the analysis of admixed genomes. Our haplotype-based implementation of the algorithm is packaged into the software *PCAmask* and details on the samples used are available in Materials and Methods and in Text S1.

*Native American Ancestral Components*

Our initial structure analysis was based on our high-density dataset (i.e., ~390K SNPs, see Table S1), thus limited to ancestral populations with available Affymetrix SNP array data (i.e., two Mesoamerican, two Andean, and three Venezuelan native populations). To explore possible relationships with additional Native American populations, we expanded our reference panel by combining our data with Illumina 650K data for 493 individuals from 52 indigenous groups from throughout the Americas [11]. Although this analysis has fewer SNPs (i.e., ~30K SNPs), it allows us to resolve within-continent population structure around the Caribbean in much greater geographic detail.

We applied the ASPCA approach described above to project the Native American segments of admixed individuals onto the full reference panel (Figure 4A, Figure S6). PC1 separates the northernmost populations, such as Canadian Northern Amerind and Na-Dene speakers, from the rest; while Mexican Pima and Central American Cabecar define the extremes of PC2 (Figure S6). Most Native American haplotypes from the admixed genomes fall along this second axis of variation, and form clearly differentiated population clusters: one cluster is shared among Colombians and most Hondurans, while another one is shared among Cubans, Dominicans, and Puerto Ricans (no Haitian haplotypes were included due to low levels of Native American ancestry). Colombians and Hondurans cluster with Chibchan-Paezan speaking groups from Colombia and Central America, including Kogi, Waunana, and Embera. In contrast, Caribbean islanders cluster with Equatorial-Tucanoan speakers, which is a major linguistic group spread across the Amazonia. With few exceptions, Equatorial-Tucanoans form a rather tight cluster, including Guahibo, Ticuna, Palikur, Karitiana, among others, many of which are settled around fluvial territories of the rainforest. This fact may have facilitated communication from and to the coast, explaining their relationship with Caribbean native components. Interestingly, the indigenous component of insular Caribbean samples seems to be shared across the different islands, indicating gene flow across the Caribbean basin in pre-Columbian times.

To explore this possibility into more detail, we performed a model-based clustering analysis using the full reference panel of 52 Native American populations from Reich et al. [11] in addition to our three native Venezuelan populations. Individual admixture proportions from K=2 through 20 are given in Figure S7. Focusing on Native American components, the first sub-continental signal (at K=4) was accounted for a Chibchan component mainly represented by the Cabecar from Costa Rica and the Bari from Venezuela. Higher order clusters pulled out Amazonian population isolates such as the Surui and Warao, as well as northern populations including the Eskimo-Aleut and Pima, in agreement with the outliers detected in our ASPCA analysis (Figure S6). Interestingly, from K=5 through 10, the Chibchan component is shared at nearly 100% with the Yukpa sample located near the Venezuelan coast, and at nearly 20% with Mayans from the Yucatan peninsula (Figure 4B). The presence of considerable proportions of the Chibchan component in the Mayan sample is indicative of possible "back" migrations from Central America and northern South America into the Yucatan peninsula, revealing an active gene flow across the Caribbean, probably following a coastal or maritime route. Moreover, very high order clusters maintain the connection between Mayans and South American components. For example, at K=16 (the model with the lowest cross validation error; Figure S8b), as much as an average of 35% of the genome in Mayans is shared with a mixed Chibchan/Equatorial-Tucanoan component mainly represented by Ticuna, Guahibo, Embera, Waunana, and Arhuaco, among others (Figure 4C). This observation is in agreement with our ASPCA results and reinforces the notion of a South American expansion of Native American components across the Caribbean basin.

*European Ancestral Components*

We performed ASPCA analysis by projecting the European segments of admixed individuals onto an extensive reference panel of European source populations, including 1,387 individuals from all over Europe sampled as part of the POPRES project [9], as well as additional Iberian samples from Galicia, Andalusia, and the Basque country in Spain (Rodriguez et al., in revision). The total projection involved 2,882 European haplotypes and 255 haplotypes of European ancestry from the admixed populations. Figure 5 shows the projection of the first two PCs where the background samples recapitulate a PCA map of Europe as reported before [15,24]. While most of the additional Iberian samples cluster together with the POPRES individuals sampled as Portuguese and Spanish, the Basques cluster separately from the centroid of most Iberian samples. The Basques are known for their historical and linguistic isolation, which could explain their genetic differentiation from the main cluster due to drift. Given the known Iberian origin of the first European settlers arriving into the Caribbean and surrounding territories of the New World, one would expect that European blocks derived from admixed Latino populations should cluster with other European haplotypes from present day Iberians. Indeed, our Latino samples aggregate in a well-defined cluster that overlaps with the cluster of samples from the Iberian Peninsula (i.e., Portugal and Spain). However, we observed that the centroid is substantially deviated with respect to the Iberian cluster (bootstrap p-value $<10^{-4}$, see Materials and Methods), suggesting the possibility of a bottleneck and drift impacting the European haplotypes of Latinos.

Importantly, when we applied ASPCA analysis using the exact same reference panel of European samples but projecting Mexican haplotypes of European ancestry (Moreno-Estrada, Gignoux et al., in preparation), we did not observe a deviated clustering pattern from the Iberian cluster: the effect is much weaker and not significant (bootstrap p-value = 0.099, see Figure S9). Furthermore, the deviation of the European segments of Mexican individuals from the distribution of the rest of Iberian samples is even smaller than the deviation of the Portuguese from the Spanish samples. We further evaluated whether the dispersion of the different subpopulations within the Caribbean cluster follow particular patterns along ASPC2, the axis driving the deviation from the Iberian centroid. We observed that Colombians and Hondurans tend to account for lower (more deviated) ASPC2 values compared to Cubans, Dominicans, and Puerto Ricans (Figure S10), suggesting a mainland versus insular population differentiation. We then performed a Wilcoxon rank test to contrast ASPC2 for mainland (Colombia and Honduras) versus island (Cuba, Dominican Republic and Puerto Rico), resulting in a highly significant p-value (1.5e-15). Because >25% of European ancestry was required for inclusion in ASPCA analysis, only two Haitian haplotypes were projected, and thus not included in the statistical analysis. Nonetheless, it is noteworthy that one of them clusters with the French, in agreement with historical and linguistic evidence about European settlements in the island (see arrow on Figure 5).

Among European populations, Iberians also have the highest proportion of identical by descent (IBD) segments that are shared with Latino populations, as measured by WELat, a statistic that is informative of the total amount of shared DNA between pairs of populations (see Figure S11 and Text S2). To explore the distribution of IBD sharing within continental groups, we considered Caribbean Latinos and Europeans separately by summing the cumulative amount of DNA shared IBD between each pair of individuals within each group. If European segments from Latino populations derive from a reduced number of European ancestors, then IBD sharing is expected to be increased among Caribbean individuals compared to European source individuals. Indeed, we observed a higher number of pairs sharing larger total IBD segment lengths among Latino individuals than among Europeans (Figure S12). Within-population endogamy is also compatible with increased IBD sharing. However, this is more likely to occur between individuals from the same subpopulation (e.g., COL-COL) rather than individuals from geographically separated subpopulations (e.g, COL-PUR). For this reason we considered Latinos as a single group as a measure to minimize such possible effect. Yet, we observed an increased proportion of IBD sharing among Latinos, arguing for a shared founder effect.

These results are in agreement with our cluster-based analysis focused on global ancestry proportions, where the European ancestry of Latinos is dominated by a shared Latino-specific component differentiated from both southern and northern European components, although shared to some extent with Spanish and Portuguese (Figure 1C). Bottlenecked populations may exhibit higher levels of differentiation from their parental gene pool due to loss of prior diversity and shift in allele frequencies. According to $F_{ST}$ estimates between $K=8$ ancestral clusters (from Figure 1C), the differentiation between southern and northern European components is 0.02. In the absence of drift, a southern-derived Latino component would be expected to show lower $F_{ST}$ values against its closely related southern component. However, the $F_{ST}$ between these two components was 0.021 (Table S3), meaning that the differentiation of the Latino-specific component with respect to southern Europeans is at least as high as the north-south differentiation within Europe. This observation was replicated when including additional Latino and ancestral populations (Figure S7). Given the increased number of divergent clusters, we focused on K=18 through 20, in which all sub-European components were jointly detected. In this case, the Latino-specific component shows further fragmentation into two components: one predominantly shared among insular Caribbean samples and the other among mainland Latinos. The $F_{ST}$ value for southern versus northern European differentiation was 0.039, while values for southern versus insular (0.041) or mainland Latinos (0.04) were slightly inflated (Table S4), supporting the notion of additional differentiation impacting the European lineages of present day admixed Latinos.

*African Ancestral Components*

The Caribbean hosts a rich history of population exchange with the African continent as a result of unprincipled slave trade practices during European colonialism. Its proximity with the North Atlantic Ocean facilitated nautical contact with the West African coast and resulted in greater exposure to slave trade routes for the local population and, ultimately, in genetic admixture. The proportion of African ancestry is consistently higher in Caribbean populations compared to individuals from the mainland (Figure 1C), and this finding is consistent across studies [3,6,25]. To explore the sub-continental composition of African segments derived from Caribbean admixed genomes, we performed ASPCA analysis on individuals with more than 25% of African ancestry using a diverse panel of African populations as potential sources (see Table S1). Our first approximation showed no dispersion of Afro-Caribbean haplotypes over PCA space. Instead, they form a relatively tight cluster that overlaps with that of the Yoruba sample from southwestern Nigeria (Figure S13). This is a plausible result, given the extensive historical record supporting a West African origin for the African lineages in the Americas.

However, according to our tract length analysis, there is strong genetic evidence for the occurrence of at least two pulses of African migrants imprinting different genomic signatures in present day admixed Caribbean populations. This poses the question of whether both pulses involved the same source population during the admixture process. If this were the case, it would easily explain our ASPCA results, where all African haplotypes point to a single source.

Alternatively, if more than one source was involved and if enough mixing occurred since the two pulses, it is possible that what we see is the midpoint of the two source populations, causing the difference to remain undetected by our standard ASPCA approach (which gives a point estimate averaging the signature of all African blocks along the genome). Hence we applied a different strategy, in which ASPCA is performed separately for short (thus older) and long (younger) ancestry tracts. For this purpose, we split the African segments of each haploid genome into two categories based on a 50cM length cutoff, and intersected the data with a reference panel of West African populations (Figure 6A). Then, for each individual, we compute assignment probabilities of coming from each of the putative parental populations based on bivariate normal distributions fitted around each PCA cluster (see Materials and Methods, Figure S14). In Figure 6B we present the scaled mean probabilities for long (>50 cM) versus short (<50 cM) African tracts in Puerto Rican individuals. The pattern that emerges reveals that African haplotypes shorter than 50 cM are more likely to have originated from populations in the coastal Northwest region, such as the Mandenka and Brong; whereas longer haplotypes show higher probabilities of coming from populations closer to the Gulf of Guinea and Equatorial West Africa, including Yoruba, Igbo, Bamoun, Fang, and Kongo (see map on Figure 6A). The significant increase in old, short Mandenka tracts when compared to longer, more recent tracts, was replicated in other insular Caribbean populations, including Cubans and Dominicans. The Brong also seem to have had a greater contribution deeper in the past not only in Puerto Ricans, but also in Dominicans, Hondurans and to a lesser extent in Colombians. In Cubans, the trend is reversed and the Brong seem to have contributed more to long tracts than to short ones (Figure S15).

     One caveat is that short ancestry tracts are more likely to be misassigned. To rule this out as a source of the signal, we added an intermediate block size category (>5 cM and <50 cM) and repeated the size-based ASPCA analysis. We observed that, despite the signal being somewhat weaker due to the lesser amount of overall data, a similar trend was retained after the exclusion of extremely short tracts (Figure S15). Finally, we gathered additional evidence by running local ancestry estimation on the African blocks alone to distinguish Mandenka vs. Yoruba ancestry tracts (see Materials and Methods). We then binned all segments of inferred Mandenka ancestry into different block sizes and observed that the proportion of the African ancestry called Mandenka is higher within shorter block sizes and decreases as block size increases (Figure 6C). This gives additional support for the differential origin of African segments and argues that such signal is not driven by the shortest genomic segments alone, but rather characterized by a progressive decay of haplotype length from older migrations as younger segments (of different ancestry) account for the majority of longer African tracts in Caribbean genomes.

**Discussion**
*Models of admixture for Caribbean and mainland populations*
     Our results reveal consistent differences regarding the admixture processes occurring across the Caribbean islands as compared to those occurring in neighboring mainland populations. First, our data suggest multiple pulses of African migration contributed significantly to genetic ancestry in the Caribbean, consistent with records of historical slave trade routes. In contrast, we find evidence of a single gene flow event of Native American ancestry into admixed Caribbean populations. Since Native American tracts are shorter, on average, than tracts of African ancestry (and therefore older), this suggests the migration event is the initial founding of admixed populations at the time of European contact. Mainland populations from Colombia and Honduras, on the other hand, are best fit by a model of repeated migration events of European

ancestry, consistent with a continuing expansion of Europeans during colonialism. We also find longer Native American tracts than African ancestry tracts in mainland populations, indicating a single pulse of the latter and a greater contribution of Native Americans into admixed continental populations. Admixture timing estimates also show consistent differences between these two groups, with admixture starting around 16-17 generations ago in the islands and 14 generations ago in mainland populations.

Our model shows remarkable agreement with historical records. The earliest European voyage by Christopher Columbus took place in 1492 (i.e., 16-17 generations ago), reaching the Caribbean island of Hispaniola (today's Dominican Republic and Haiti). Later European voyages reached the coasts of Central and South America, so permanent European settlements did not occur in the mainland until the first half of the XVI century, consistent with an approximate difference of 2 generations between the estimated onset of admixture according to our island and mainland models. Here we have focused on Colombians and Hondurans as population samples from mainland territories with coastal access into the Caribbean, but we have previously reported admixture timing estimates for Mexicans as well, namely starting 15 generations ago [5]. The settlement of Europeans in mainland Mexican territory is documented to have occurred between 1519 and 1521 (i.e., 27-29 years apart from the first contact in 1492 in the Caribbean), that is ~1 generation apart between the average estimate of 16 generations for the onset of admixture in the Caribbean compared to 15 generations from our model based on Mexican data. The abundance of historical records about European colonization of the New World is particularly exceptional, facilitating the contrast between written and genetic registries.

*South American origin of indigenous components in the Caribbean*
In contrast with other regions in the Americas where indigenous peoples are numerous, the genetic characterization of Native American components in the Caribbean required indirect reconstruction via genomic assembly of indigenous ancestry tracts transmitted into extant admixed individuals. By applying ancestry-specific PCA and cluster-based analyses integrating a large number of indigenous groups throughout the Americas, we found that Equatorial-Tucanoan speakers from South America hold the closest relationship with Caribbean indigenous components. This was also observed in a different sample set from the 1000 Genomes Project (Gravel et al., submitted). Despite covering a large geographic area of South America (ranging from northern Colombia and Brazil to southern Bolivia and Paraguay), most Equatorial-Tucanoans cluster together in PCA space, arguing for a common origin. We have intentionally included three additional tribes from the Venezuelan coast since logical candidates for the origin of the ancestors of Caribbean populations include indigenous coastal groups south the Lesser Antilles. However, despite their closer geographic location, none of these groups primarily accounted for the indigenous ancestry of the insular Caribbean samples, pointing to an inland origin rather than a coastal one. Nonetheless, our cluster-based analysis revealed that native Venezuelan components do share membership with several Central American indigenous populations, such as the Costa Rican Cabecar, and, to a lesser extent, with Mayan groups from the Yucatan peninsula of present day Mexico, suggesting substantial gene flow across the Caribbean Sea in pre-Columbian times. In fact, based on the distribution of jade, obsidian, pottery, and other commodities, archaeological evidence supports the existence of maritime-based interaction networks between central Mesoamerica, the Isthmo-Colombian area, and northern Venezuela [26]. Our results demonstrate that such long distance negotiations were accompanied by gene exchange between previously diverged native populations, and give a richer perspective of the dynamics between the inhabitants of the Caribbean basin prior to European contact.

In a recent genomic survey of the relationships between Native American peoples, Reich and colleagues [11] described the Chibchan speakers on both sides of the Panama isthmus as an exception to the simple model involving a southward expansion with sequential population splits and little subsequent gene flow. Instead, Central Americans, such as the Cabecar from Costa Rica, were modeled as a mixture of South and North American ancestry, which the authors reported as evidence for a back-migration from South into Central America. Our findings not only support such interpretation, but also suggest a distant connection between Caribbean Mesoamerica and South American inland territories. Specifically, the fact that Mayans from the Yucatan peninsula share 35% of their genome with Ticuna, Guahibo, and Arhuaco, among other Chibchan and Equatorial-Tucanoan speakers, supports the expansion of an inland South American component across the Caribbean. For context, it is noteworthy that in ASPCA the native ancestry tracts of Colombians and Hondurans cluster with geographically closer indigenous tribes, such as Chibchan speakers from Colombia and Central America. How to account, then, for a shared clustering between more distant Equatorial-Tucanoan speakers (mostly of Amazonian origin), and insular Caribbean haplotypes? One possible explanation is that the fluvial nature of most of these settlements (across the Amazon and Orinoco basins) may have facilitated people movement to the coast, and eventually migrating north through the Lesser Antilles, explaining their relationship with Caribbean native components. In fact, our results are consistent with archaeological records suggesting that the ancestors of the indigenous people that Columbus encountered might have come from populations that migrated from the Lower Orinoco Valley around 3,000 years ago [27,28]. Additionally, our results align with the classification of languages spoken by pre-Columbian inhabitants of the Caribbean. Together with Caribs, Tainos were the major group living in the Greater Antilles and surrounding islands at the moment of European contact. Tainos and insular Caribs spoke Arawakan languages that belong to the Equatorial sub-family, in the Equatorial-Tucanoan family [29]. The geographic distribution of Arawakan languages across northern South America resembles the sampling coverage of the Equatorial-Tucanoan individuals analyzed here (see map in [11]), supporting our findings. Previous genetic studies have also pointed to a South American origin for Tainos [7,30]. Based on mitochondrial haplogroups ascertained from pre-Columbian human remains, Lalueza-Fox and colleagues [30] found that only two of the major mtDNA lineages, namely C and D, were present in their sample (N=27). Given that high frequencies of C and D haplogroups are more common in South American populations, the authors argued for that sub-continent as the homeland of the Taino ancestors.

Overall, our analysis of indigenous ancestry tracts from extant admixed genomes supports previous linguistic, archaeological, and ancient DNA evidence about the peopling of the Caribbean, and goes beyond by pointing to a greater involvement of inland Amazonian populations during the last migration into the Antilles prior to European contact. Earlier migrations may have occurred (e.g., from Mesoamerica or the Florida peninsula), as pre-ceramic archaeological evidence of human presence in the Greater Antilles dates back more than 7,000 years ago [27]. However, the fact that the Equatorial-Tucanoan component is shared among the indigenous haplotypes from different insular and continental populations supports either a single South American origin of Caribbean settlers or a major population replacement involving a more recent migration of agriculturalists from inland South America.

*Founder effect in the European lineage of admixed Latinos*

We find genomic patterns compatible with the effect of a founder event in the ancestral European population of present day admixed Latinos. Supporting evidence include: 1) a Latino-specific European component revealed by clustering algorithms, which is not assigned to source populations within Europe except Spain and Portugal, and detected at lower order clusters compared to other European and Native American sub-continental components; 2) inflated $F_{ST}$ values between the Latino-specific and southern European components, compared to southern versus northern Europe differentiation; 3) significant deviation of the distribution of European haplotypes from the main cluster of Iberian samples in ASPCA space; and 4) increased IBD sharing among Latino individuals compared to Europeans. Additionally, a similar signature was observed in an independent dataset of Latino samples from the 1000 Genomes Project using a combined approach that integrates IBD and local ancestry tracts (Gravel et al., submitted). These findings suggest that early European waves of migration into the New World involved a reduced ancestral population size, mainly composed by Iberians, bearing a subset of the diversity present within the source population, causing the derived admixed populations to diverge from current European populations. Furthermore, we find differences between mainland and insular Caribbean populations including 1) different time estimates for the onset of admixture as revealed by ancestry tract length analysis (Figure 3); 2) separate memberships in cluster-based analyses (Figure 4C, Figure S7); and 3) significantly shifted distributions within the Latino cluster in ASPCA projection of European haplotypes (Figure 5, Figure S10). The fact that mainland Colombians and Hondurans show not only the highest proportions of the Latino-specific European component in ADMIXTURE but also the most extreme deviation from the Iberian cluster in ASPCA, suggests stronger genetic drift in these populations, compatible with a two-stage European settlement involving insular territories at first, and mainland populations subsequently absorbing a subset of migrants from the islands.

There is documented evidence of extensive migration from the islands to the continent throughout the 16[th] century [21]. There were only two viceroyalties of the Spanish Empire in the New World until the 18[th] century –Viceroyalty of New Spain (capital, Mexico City) and Peru (capital, Lima)–. An additional viceroyalty in South America was created in 1717 with Bogota as capital (Viceroyalty of New Granada), promoting economic and population growth. Interestingly, the estimated time for the second pulse or European migrants into the admixture of present day Colombians (i.e., 12 generations ago) coincides with the creation of the Colombian-based Viceroyalty of New Granada, accounting for the large increase (from 12.5% to 75%) of European ancestry in the model based on tract length distributions. Such small contribution of European ancestry at the onset of admixture in Colombians reinforces the idea that their patterns of European diversity are heavily impacted by a reduced number of founders. In contrast, Mexican-derived European haplotypes do not appear to be impacted by founder events as much as the Caribbean populations analyzed here. A possible explanation is that present day Mexico was the center of the wealthy Viceroyalty of New Spain, constituting one of the largest European settlements under Spanish rule, ensuring continuous exchange with Spain throughout colonial times, resulting in a larger ancestral population size.

*Space and time distinction of African migrations into the Caribbean*

We find that populations from the insular Caribbean are best modeled as mixtures absorbing two independent waves of African migrants. Assuming a 30-year generation time [22], the estimated average of 15 generations ago for the first pulse (i.e., circa 1550) agrees with the introduction of African slaves soon after European contact in the New World. At first, local natives were used as the source of forced labor, but populations were decimated rapidly, giving rise to the four century long transatlantic slave trade, which is usually divided into two eras. The first one accounted for a small proportion (3-16%) of all Atlantic slave trade, whereas the second Atlantic system peaked in the last two decades of the 18$^{th}$ century, accounting for more than half of the slave trade. This period of increased activity coincides with the estimated age of the second (and stronger) pulse of African tracts according to our model (e.g., 7 generations ago in Dominicans, one of the major absorbing populations, pointing to late 18$^{th}$ century). In other words, the estimated time separation between these two pulses (i.e., 8 generations or ~240 years) based on genetic data is in extraordinary agreement with historical records, recapitulating the span between the onset of African slave trade and its period of maximum intensity right before its rapid decline during the 19$^{th}$ century [31].

To address the question of whether there was also a separation in space between the origins of these two pulses, we relied on the fact that chromosomes from older contributions to admixture have undergone more recombination events, thus leading to shorter continuous African ancestry tracts. By conducting two different but complementary size-based analyses restricted to genomic segments of inferred African ancestry, we provide compelling evidence that short African tracts are enriched with haplotypes from northern coastal West Africa, represented by Mandenka samples from Senegal and Brong from western Ghana, near the Ivory Coast. This is in agreement with documented deportation flows during the 15$^{th}$-16$^{th}$ centuries, where most enslaved Africans were carried off from Senegambia and departed for the Americas from the Gorée Island, near Cape Verde [31]. African slaves were embarked by European traders in ports along the West African coast, but raiding zones extended inland with the involvement of local African kingdoms. The Mandinka Kingdom of Senegambia was part of the Mali Empire, one of the most influential domains in West Africa, spreading its language, laws, and culture along the Niger River. The empire's total area included nearly all the land between the Sahara Desert and coastal forests, and by 1530 reached modern-day Ivory Coast and Ghana, possibly accounting for the shared pattern between Mandenka and Brong with respect to Caribbean's short ancestry tracts. While such interpretation is supported by the fact that Mandenka and Brong are the westernmost population samples of our reference panel, the lack of additional samples from northern West Africa prevent us from determining whether this pattern is shared with other tribes as well. On the other hand, the greater affinity of longer ancestry tracts with the rest of the samples, which cover much of the central West African coast, is compatible with the greater involvement of such regions in the slave trade during the 18$^{th}$ century.

The volume of captives being embarked from the bights of Benin (e.g., today's Nigeria) and Biafra (e.g., today's Cameroon) was so elevated after 1700 that part of its shore soon became known as the "Slave Coast" [31]. Population samples around this area represented in our reference panel include Yoruba and Igbo from Nigeria, and Bamoun and Fang from Cameroon, all of which show higher probabilities of being assigned as the source for longer African ancestry tracts in the admixed Latino groups analyzed. In fact, together with Brazil, the Caribbean Islands were the major slave import zone during the 18$^{th}$ century. Later deportation flows in the 19$^{th}$ century involved ports of origin near the Congo River in West Central Africa. The closest population sample of our reference panel from this region is represented by the Kongo, which also shows higher affinity with longer ancestry tracts, compatible with a later contribution into the admixture of the Caribbean. The 19$^{th}$ century also saw the abolition of slavery in most parts of the world, but the massive international flow of people it involved, remains as one of the deepest signatures in the genome of descendent populations. While the geographic extension of the regions of origin of African slaves brought to the Americas has been widely documented, it was unclear until now the extent to which particular sub-continental components have shaped the genomic composition of present day Afro-Caribbean descendants. Our ancestry-specific and size-based analyses allowed us to discover that African haplotypes derived from Caribbean populations still retain a signature from the first African ancestors despite the later dominance of African influx from multiple sub-continental components.

*Conclusion*

Our genome-wide dense genotyping data from six different populations of Caribbean descent, coupled with the availability of large-scale reference panels, allowed us to address long-standing questions regarding the origin and admixture history of the Caribbean Basin. The differences between insular and continental Caribbean populations underscore the importance of characterizing admixed populations at finer scales. We report ancestry-specific recent bottlenecks affecting particular Latino groups, but not others, which may have important implications in the expected relative proportion of deleterious mutations and elevated allele frequencies that can be detected via association studies in theses populations. Finally, the extensive population stratification within sub-continental components implies that medically relevant genetic variants may be geographically restricted, which reinforces the need for sequencing target populations in order to discover local variants that may only be relevant in Latino-specific association studies for disease.

**Materials and Methods**

*Samples and Data Generation*

Generated data and assembled datasets for this study are summarized in Table S1. A total of 251 individuals representing six different Caribbean-descent populations were recruited in South Florida, USA. Participants were required to have at least three grandparents from their countries of origin, thus limited ethnographic and anonymous pedigree information was collected. The majority of pedigrees (94.3%, n=82) had four grandparents from the same country. Only 5 pedigrees (5.7%) had one grandparent from a different country. Informed consent was obtained from all participants under approval by the University of Miami Institutional Review Board. A total of 76 trios, 2 duos, and 19 parents were genotyped using Affymetrix 6.0 SNP arrays, which included: 80 Cubans, 85 Colombians, 34 Dominicans, 27 Puerto Ricans, 19 Hondurans, and 6 Haitians. Out of 173 founders, 18 samples were filtered from structure analyses due to cryptic relatedness as inferred by IBD>10%. Four trios were not considered for trio phasing due to an excess of Mendelian errors (>100K), two trios were removed due to $3^{rd}$ or higher degree of relatedness between parents as inferred by IBD, and five trios were filtered due to cryptic relatedness between members of different trios above 10% IBD. After filtering, 65 complete trios remained for haplotype-based analyses. To study population structure and demographic patterns involving relevant ancestral populations, 79 previously collected samples from three native Venezuelan tribes were genotyped using the same array (i.e., 25 Yukpa [aka Yucpa], 29 Bari, and 25 Warao). We combined our data with publicly available genomic resources and assembled a global database incorporating genome-wide SNP array data for 3,042 individuals from which two datasets with different SNP densities were constructed (see Table S1). The high-density dataset included populations with available SNP data from Affymetrix arrays; namely African, European, and Mexican HapMap samples [8], Europeans from POPRES [9], West Africans from Bryc et al. [10], and Native Americans from Mao et al. [32]. After merging and quality control filtering, 389,225 SNPs remained and representative population subsets were used in different analyses as detailed through sections below. Our lower density dataset (30,860 SNPs) resulted from the intersection of our high-density dataset with available SNP data generated on Illumina platform arrays, including 52 additional Native American populations [11], as well as additional Latino populations sampled in New York City [7] and 1000 Genomes Latino samples [6]. The resulting dataset combines genomic data for 1,262 individuals from 80 populations. Full details on the population samples are available in Table S1.

*Population Structure*

An unsupervised clustering algorithm, ADMIXTURE [12], was run on our high-density dataset to explore global patterns of population structure among a representative subset of 641 samples, including seven Native American, eleven PopRes European, HapMap3 Nigerian Yoruba, HapMap3 Mexican, and our six new Caribbean Latino populations (see Table S1). Fourteen ancestral clusters (*K*=2 through 15) were successively tested. Log likelihoods and cross-validation errors for each *K* clusters are available in Figure S3. $F_{ST}$ based on allele frequencies was calculated in ADMIXTURE v1.22 for each identified cluster at *K*=8 and values are available in Table S3. Our low-density dataset comprising 1,262 samples (detailed in Table S1) was used to run *K*=2 through 20. Log likelihoods, cross validation errors and $F_{ST}$ values from ADMIXTURE are available in Figure S8 and Table S4. Principal component analysis (PCA) was applied to both datasets using EIGENSOFT 4.2 [33] and plots were generated using R 2.15.1. Global ancestry estimates from ADMIXTURE at *K*=3 were used to test the correlation between male and female ancestry proportions considering all trio founders within each Caribbean population as well as within the full set of admixed trios. Linear models and permutations (up to 100,000) were performed using R 2.15.1.

*Phasing and Local ancestry assignment*

Family trio genotypes from our six Caribbean populations and continental reference samples were phased using BEAGLE 3.0 software [34]. Local ancestry assignment was performed using PCAdmix (http://sites. google.com/site/pcadmix/ [19]) at *K*=3 ancestral groups. This approach relies on phased data from reference panels and the admixed individuals. To maintain SNP density and maximize phasing accuracy we restricted to a subset of reference samples with available Affymetrix 6.0 trio data, namely 10 YRI, 10 CEU HapMap3 trios, and 10 Native American trios from Kidd et al. [5]. Each chromosome is analyzed independently, and local ancestry assignment is based on loadings from Principal Components Analysis of the three putative ancestral population panels. The scores from the first two PCs were calculated in windows of 70 SNPs for each panel individual (in previous work we have estimated a suitable number of 10,000 windows to break the genome into when inferring local ancestry using PCAdmix, and in this case, after merging Affymetrix 6.0 data from admixed and reference panels, a total of 743,735 SNPs remained /10,000 = window length of ~70 SNPs). For each window, the distribution of individual scores within a population is modeled by fitting a multivariate normal distribution. Given an admixed chromosome, these distributions are used to compute likelihoods of belonging to each panel. These scores are then analyzed in a Hidden Markov Model with transition probabilities as in Bryc et al. [10]. The g (generations) parameter in the HMM transition model was determined iteratively so as to maximize the total likelihood of each analyzed population. Local ancestry assignments were determined using a 0.9 posterior probability threshold for each window using the forward-background algorithm. In analyses that required estimating the length of continuous ancestry tracts, the Viterbi algorithm was used. An assessment of the accuracy of this approach is given in [5].

*Tract length analysis*

We used the software *Tracts* [20] to identify the migratory model that best explains the genome-wide distribution of ancestry patterns. Specifically, we considered three migration models, each featuring a panmictic population absorbing migrants from three source populations. The models differ by the number of allowed migration events per population. In the simplest model, the population is founded by Native American and European individuals, and later receives a pulse of African migrants. The initial ancestry proportion and timing, as well as the African migration amplitude and timing, are fitted to the data as described below. The other two models feature an additional input of either European or African migrants; the timing and magnitude of this additional pulse result in two additional parameters that must be fitted to the data. Here, the data consisted of Viterbi calls from PCAdmix (see previous section and Figure 2), that is, the most probable assignment of local ancestry along the genomes. To fit parameters to these data, we tallied the inferred continuous ancestry tracts according to inferred ancestry and tract length using 50 equally spaced length bins per population, and one additional bin to account for full chromosomes. Given a migration model and parameters, *Tracts* calculates the expected counts per bin. Assuming that counts in each bin are Poisson distributed, it produces a likelihood estimate that is used to fit model parameters. For each population, we report the model with the best Bayesian Information Criterion (BIC) $-2 \log(L) + k \log(n)$, with $n=153$.

*Ancestry-Specific Principal Component Analysis (ASPCA)*

To explore within-continent population structure, we applied the following approach for each of the continental ancestries (i.e., Native American, European, and African) of admixed genomes. The general framework is shown in Figure 2. It comprises locus-specific continental ancestry estimation along the genome, followed by PCA analysis restricted to ancestry-specific portions of the genome projected onto sub-continental reference panels of ancestral populations. For this purpose, we used our continental-level local ancestry estimates provided by PCAdmix to partition each genome into ancestral haplotype segments, and retained for subsequent analyses only those haplotypes assigned to the continental ancestry of interest. This is achieved by masking (i.e., setting to missing) all segments from the other two continental ancestries. Because ancestry-specific segments may cover different loci from one individual to another, a large amount of missing data results from scaling this approach to a population level, which limits the resolution of PCA. To overcome this problem, we adapted the subspace PCA (ssPCA) algorithm introduced by Raiko et al. [35] to implement a novel ancestry-specific PCA (ASPCA) that allows accommodating phased haploid genomes with large amounts of missing data. Our method is analogous to the ssPCA implementation by Johnson et al. [23], which operates on genotype data. In contrast, ASPCA operates on haplotypes, allowing us to use much more of the genome (rather than just the parts estimated to have two copies of a certain ancestry) and to project independently the two haploid genomes of each individual. Finally, ancestry-specific haplotypes derived from admixed individuals are combined with haplotypes derived from putative parental populations and projected together onto PCA space. Details of the ASPCA algorithm and constructed datasets are described in Text S1.

*Differentiation of sub-European ancestry components*

To measure the observed deviation in ASPCA of European haplotypes derived from admixed Caribbean populations with respect to the cluster of Iberian samples, a bootstrap resampling-based test was performed. The null distribution was generated from comparing bootstraps of Portuguese and Spanish ASPCA values as models of the intrinsic Iberian population structure. We then compared the ASPCA values of the admixed individuals and tested if the observed differences between Iberian ASPCA values and those of the admixed individuals are more extreme than the differences within Iberia. The distance was determined using the chi-squared statistic of Fisher's method combining ASPC1 and ASPC2 t-tests for each bootstrap. We ran 10,000 bootstraps to determine one-tailed p-values. As Iberians we considered: PopRes Spanish, PopRes Portuguese, Andalusians, and Galicians; and as Caribbean Latinos: CUB, PUR, DOM, COL, and HON. Additional tests were performed comparing Portuguese versus the rest of Iberians and between an independent dataset of Mexican individuals analyzed by Moreno-Estrada, Gignoux et al. (in preparation) projected onto ASPCA space using the same reference panel of European populations. A bivariate test was performed to measure the relative deviation from the Iberian cluster of the distribution given by the Caribbean versus the Mexican dataset. To determine whether insular versus mainland Caribbean populations disperse over significantly different ranges in ASPC2, a Wilcoxon rank test was performed between (COL+HON) versus (CUB, PUR, DOM). Haitians were excluded due to low sample size (N=2 haplotypes). Boxplot is available in Figure S10. Population differentiation estimates between clusters inferred with ADMIXTURE were visualized and compared across runs where both the Latino-specific and southern European components were detected. Values are available in Table S3 and Table S4. The analysis of IBD sharing was conducted using our high-density dataset extracting a subset of 203 PopRes European individuals and the founders from the 65 complete admixed trios. We first performed a genome-wide pairwise IBS estimation using PLINK [36] to ensure that the dataset contains no samples with more than 10% IBS with any other sample. Then we used fastIBD [34] to phase the data and estimate segments shared IBD longer than 2 Mb to eliminate false positive IBD matches and assuming that ancestry will be shared among pairwise IBD hits of segments this long. All 2 Mb or greater segments shared IBD between pairs of individuals were summed, and histograms were created for pairwise matches within each group (i.e., PopRes Europeans and Caribbean Latinos).

*Size-based ASPCA analyses*

Given the evidence from our tract length analysis for a second pulse of African migrants into the admixture of insular Caribbean Latinos, a modified size-based ASPCA analysis was performed. A reference panel was built integrating three different resources [8,10,37] and focusing on putative source populations from along the West African coast, including Mandenka from Senegal, Yoruba and Igbo from Nigeria, Bamoun and Fang from Cameroon, Brong from western Ghana, and Kongo from the Democratic Republic of the Congo. We begin with the continental local ancestry inference from PCAdmix *K*=3. For each individual we then divide African ancestry tracts into small (0 to 50 cM) and large (> 50 cM) size classes. Given a partition of African ancestry tracts, we take all sites included in one tract class, say short tracts, and run PCA on our sub-continental West African reference populations for only these sites. Using the first two PCs from this analysis, we fit a bivariate normal distribution to each reference population cluster. We then project our test sample into this PCA space, and estimate the probability of it coming from each reference population using the fitted distributions. This procedure is repeated for each tract class, for each individual. For each admixed Caribbean population, we can then estimate the probability that a given class of African ancestry tracts comes from a specific West African source population as the average probability of assignment to this population across all individuals. Finally, under the assumption that a given class of African tracts must come from one of the provided reference populations, we rescale these probabilities to sum to one. Each assignment estimate is also provided with error bars representing the standard error of the mean. We compare the short and long assignment probabilities for each Caribbean population to identify distinct sources for "older" and "younger" West African migratory source populations. Haitians were not included in the analysis due to low sample size (n=4). Due to concerns that shorter tracts have a higher likelihood of mis-assignment, we added a medium tract size class (5cM to 50 cM) to see if the results were simply due to very short (0 cM to 5 cM) European or Native American tracts being mis-classified as African. We compare the results for short and medium tracts and find that the trends are maintained suggesting the observation that older shorter tracts appear to be primarily from the Mandenka and Brong source populations is not simply due to short tract mis-assignment

*Local ancestry estimation within African tracts*

To identify likely regions of Yoruba versus Mandenka ancestry in the African component, we modified our implementation of PCAdmix to perform local ancestry deconvolution solely of the African segments of the admixed genomes. The modification is achieved in the final step of the algorithm: whereas the standard approach estimates a single HMM across an entire chromosome, here we fit J disjoint HMMs spanning each of the J blocks of African ancestry in a given chromosome for a given individual. Applying the method, we obtained posterior probabilities for Mandenka versus Yoruba ancestry within the previously inferred African segments. We then selected only those sub-regions that were confidently called as Mandenka or Yoruba, and stratified them by physical size.

**Acknowledgements**
We thank study participants for generously donating DNA samples, Brenna M. Henn, Martin Sikora, and Meredith Carpenter for helpful comments on earlier versions of the manuscript, Scott Huntsman for data management, and David Reich and Andres Ruiz-Linares for data sharing. This project was supported by NIH grant 1R01GM090087 to ERM and CDB, as well as P60MD006902 from the National Institute on Minority Health and Health Disparities to EGB and NIH Training Grant T32 GM007175 to CRG. This project was also supported in part by an award from the Stanley J. Glaser Foundation to JLM and ERM and by the George Rosenkranz Prize for Health Care Research in Developing Countries awarded to AM. The collections and methods for the Population Reference Sample (POPRES) are described by Nelson et al. (2008). The datasets used for the analyses described in this manuscript were obtained from dbGaP at http://www.ncbi.nlm.nih.gov/projects/gap/cgibin/study.cgi?study_id=phs000145.v1.p1 through dbGaP accession number phs000145.v1.p1. The Native American dataset from Reich et al. (2012) was obtained from the University of Antioquia through a data access agreement dated July 26, 2012.

**Figure Legends**

**Figure 1: Population structure of Caribbean and neighboring populations.** A) On the map, areas in red indicate countries of origin of newly genotyped admixed population samples and blue circles indicate new Venezuelan (underlined) and other previously published Native American samples. B) Principal Component Analysis and C) *ADMIXTURE* [12] clustering analysis using the high-density dataset containing approximately 390K autosomal SNP loci in common across admixed and reference panel populations. Unsupervised models assuming K= 3 and K=8 ancestral clusters are shown. At K=3, Caribbean admixed populations show extensive variation in continental ancestry proportions among and within groups. At K=8, sub-continental components show differential proportions in recently admixed individuals. A Latino-specific European component accounts for the majority of the European ancestry among Caribbean Latinos and is exclusively shared with Iberian populations within Europe. Notably, this component is different from the two main gradients of ancestry differentiating southern from northern Europeans. Native Venezuelan components are present in higher proportions in admixed Colombians, Hondurans, and native Mayans.

**Figure 2: Diagram of the analytical strategy used for reconstructing migration history and sub-continental ancestry in admixed genomes**. The starting point consists of genome-wide SNP data from family trios. Unrelated individuals are used to estimate global ancestry proportions with *ADMIXTURE*, whereas full trios are selected for *BEAGLE* phasing and PCA-based local ancestry estimation using continental reference samples. From here, two orthogonal analyses are performed: 1) Ancestry-specific regions of the genome are masked to separately project European, African, and Native American haplotypes onto the PCA space defined by large sub-continental reference panels of putative ancestral populations. We refer to this methodology as ancestry-specific PCA (*ASPCA*) and the code is packaged into the software *PCAmask*. 2) Continental-level local ancestry calls are used to estimate the tract length distribution per ancestry and population, which is then leveraged to test different demographic models of migration using *Tracts* software.

**Figure 3: Demographic reconstruction since the onset of admixture in the Caribbean.** We used the length distribution of ancestry tracts within each population from A) insular and B) mainland Caribbean countries of origin. Scatter data points represent the observed distribution of ancestry tracts, and solid-colored lines represent the distribution from the model, with shaded areas indicating 68.3% confidence intervals. We used Markov models implemented in *Tracts* to test different demographic models for best fitting the observed data. Insular populations are best modeled when allowing for a second pulse of African ancestry, and mainland populations when a second pulse of European ancestry is allowed. Admixture time estimates (in number of generations ago), migration events, volume of migrants, and ancestry proportions over time are given for each population under the best-fitting model. The estimated age for the onset of admixture among insular populations is consistently older (i.e., 16-17) compared to that among mainland populations (i.e., 14).

**Figure 4: Sub-continental origin of Native American components in the Caribbean.** A) Ancestry-specific PCA analysis showing the projection of haploid genomes with >3% Native American ancestry from admixed individuals (colored circles) onto a reference panel of 52

Native American populations (gray symbols) from [11] sampled throughout the continent. B) Shared indigenous components across the Caribbean as revealed by cluster-based ADMIXTURE analysis at $K$=10 of Native American populations from [11], plus three additional Native Venezuelan tribes genotyped for this project. The plot includes individuals from Mesoamerica southwards only (Eskimo-Aleut, Na-Dene, and Canadian North Amerind not shown). Legends on top correspond to linguistic families for samples from [11], and to tribe names for the three additional Venezuelan populations. Dashed gray rectangles indicate individuals included in the enlarged insets, and numbers are used to indicate their approximate geographic location. Arrows indicate bidirectional gene flow between populations. C) ADMIXTURE model for $K$=16 ancestral clusters considering the same set of samples as in B). At higher-order clusters, a South American component (in green) represented by Ticuna individuals and other Equatorial-Tucanoan groups accounts for the Mayan mixture, supporting pre-Columbian back migrations across the Caribbean.

**Figure 5: Sub-continental origin of European haplotypes derived from admixed genomes.** Haploid genomes with >25% European ancestry derived from insular Caribbean (black symbols) and mainland populations (gray symbols) are projected onto a reference panel (colored labels) of 1,387 PopRes European samples with four grandparents from the same country [15], and 54 additional Iberian individuals sampled in Spain (Rodriguez et al. in revision). PC1 values have been inverted and axes rotated 16 degrees counterclockwise to approximate the geographic orientation of population samples over Europe. Population codes are detailed in Table S1 and regions within Europe are labeled as in [16]. Inset map: countries of origin for PopRes samples color-coded by region (areas not sampled in gray and Switzerland in intermediate shade of green to denote shared membership with EUR W, EUR C, and EUR S). Most Latino-derived European haplotypes cluster around the Iberian cluster. One of the two Haitian individuals included in the analysis clustered with French speaking Europeans (black arrow), in agreement with the colonial history of Haiti and illustrating the fine-scale resolution of our ASPCA approach.

**Figure 6: Sub-continental origin of Afro-Caribbean haplotypes of different sizes**. A) Map of West Africa showing locations of reference panel populations. Samples in black are more likely to represent the origin of short ancestry tracts and those in red of long ancestry tracts, according to B) assignment probabilities for each putative ancestral population of being the source for short (<50 cM in black) and long (>50 cM in red) ancestry tracts. African ancestry tracts for Puerto Ricans are shown and results for all populations are available in Figure S15. C) Proportion of African ancestry of inferred Mandenka origin as a function of block size in the combined set of Caribbean genomes. By running *PCAdmix* within the previously inferred African segments, we obtained posterior probabilities for Mandenka versus Yoruba ancestry. Overall, we found evidence for a differential origin of the African lineages in present day Afro-Caribbean genomes, with shorter (and thus older) ancestry tracts tracing back to Far West Africa (represented by Mandenka and Brong), and longer tracts (and thus younger) tracing back to Central West Africa.

**Supporting Information: Figure Legends**

**Figure S1**: Principal component 1 versus lower order PCs defining sub-continental components among Native American populations. Top: PC5 separates Venezuelan population isolates from the rest of Native Americans. Bottom: PC7 separates Mesoamerican from Andean groups. Mexicans and Hondurans distribute between the European and Mesoamerican clusters, whereas Colombians slightly deviate towards the Andean and Venezuelan clusters. Global PCA analysis based on the high-density dataset (~390K SNPs) and thus limited to reference panel populations with available Affymetrix SNP array data (see Table S1 for details).

**Figure S2**: ADMIXTURE results from *K*=2 through 15 based on the high-density dataset (~390K SNPs) including 7 admixed Latino populations and 19 reference populations. A low-frequency Southern European component restricted to Mediterranean populations at lower order Ks and specifically to Iberian populations at higher order Ks, accounts for the majority of European ancestry among Latinos (black bars). It further decomposes into population-specific clusters (purple bars) denoting higher similarities within the European portion among Latinos compared to European source populations.

**Figure S3**: ADMIXTURE metrics at increasing K values based on Log-likelihoods (A) and cross-validation errors (B) for results shown in Figure S2.

**Figure S4**: Correlation between male and female continental ancestries. Parents' ancestry proportions from each trio were used to compare correlation coefficients between the observed values and 100,000 permuted male-female pairs (p-values shown for the combined set of Latino Caribbean samples and for each population in Table S2).

**Figure S5**: Ancestry tract lengths distribution per population and demographic model tested in *Tracts*. For each demographic scenario, the observed distribution is compared to the predictions of the best-fitting migration model (displayed below each distribution). Solid lines represent model predictions and shaded areas are one-sigma confidence region surrounding the predictions. Three different demographic scenarios were considered, all of which assume the involvement of European and Native American tracts at the onset of admixture, followed by the introduction of African migrants (denoted by *EUR,NAT +AFR*). The second and third models allow for an additional pulse of European (*EUR,NAT +AFR +EUR*) and African (*EUR,NAT +AFR +AFR*) ancestry, respectively. Likelihood values for each model are shown on top of each plot. Pie charts above each migration model are proportional to the estimated number of migrants being introduced at each point in time (black arrows). GA: generations ago.

**Figure S6**: ASPCA projection of Native American haplotypes derived from admixed genomes (solid circles) onto reference panel populations from [11] grouped by linguistic families. Top panels: ASPCA with the full reference panel of Native American populations. Bottom panels: zoomed ASPCA without extreme outliers (Aleutians, Greenlanders, and Surui excluded from the analysis).

**Figure S7**: ADMIXTURE results from *K*=2 through 20 based on the low-density dataset (~30K SNPs) including additional admixed Latino and Native American reference populations (see

Table S1 for details). The presence of the Latino European component (black and gray bars) is recaptured among independently sampled Latino populations. FL: Florida (this study); NY: New York; 1KG: 1000 Genomes Project samples. Throughout lower and higher order Ks, several South American components represented by Equatorial-Tucanoan, Ge-Pano-Carib, and to a lesser extent Chibchan-Paezan speakers (yellow and green bars), show varying degrees of shared genetic membership with Northern Amerind Mayans, accounting for up to nearly half of their genome composition (see Figure 4 for more details).

**Figure S8**: ADMIXTURE metrics at increasing K values based on Log-likelihoods (A) and cross-validation errors (B) for results shown in Figure S7.

**Figure S9**: ASPCA distribution of Iberian samples (red circles) compared to European haplotypes derived from our Latino Caribbean samples (top panel) and from an independent cohort of Mexican samples (bottom panel). The relative deviation from the Iberian cluster is significantly different comparing the Caribbean versus the Mexican dataset (see the main text for details).

**Figure S10**: ASPC2 values per population from the European-specific haplotype projection shown in Figure 5 and Figure S9. Population codes as in Table S1. The boxplot shows that low ASPC2 values are enriched with mainland Colombian and Honduran haplotypes, whereas insular Caribbean populations show less deviated values from the Iberian cluster. A Wilcoxon rank test between mainland (COL, HON) versus insular samples (CUB, PUR, DOM) demonstrated that these two groups disperse over significantly different ranges in ASPC2 (Haitians excluded due to low sample size).

**Figure S11**: Pairwise IBD sharing between Caribbean Latinos and a representative subset of PopRes European populations as measured by WELat. For each Latino population, pairwise IBD values reach maximum levels among pairs involving Portuguese and Spanish samples. See the main text for details.

**Figure S12**: IBD sharing between pairs of individuals within Caribbean Latinos (A) and a representative subset of PopRes European populations (B). Inset histograms display counts lower than 50 for the same binning categories. The overall count of pairs sharing short segments of total IBD is higher among Europeans, probably as a result of an older shared pool of source haplotypes. In contrast, the higher frequency of longer IBD matches among Latinos is compatible with a recent European founder effect.

**Figure S13**: ASPCA projection of African haplotypes derived from admixed genomes with >25% of African ancestry (black symbols) onto a representative subset of African HapMap3 and other West African reference panel populations from [10]. Colombians and Hondurans excluded due to lower overall proportions of African ancestry.

**Figure S14**: ASPCA projection of short versus long African ancestry tracts onto West African populations reference panel. To exemplify our size-based ASPCA approach, the African genome of a Puerto Rican individual is displayed (denoted by PUR). Left: PUR clusters with Mandenka when only sites within short ancestry tracts (<50 cM) are considered to perform PCA. Right: a similar background distribution is obtained but the same PUR individual no longer clusters with Mandenka when considering long ancestry tracts (>50 cM).

**Figure S15**: African ancestry size-based ASPCA results per population sample. Considering three different classes of ancestry tract lengths (black: short; red: long; blue: intermediate), scaled assignment probabilities are shown for each African source population. Values on the y-axis are the average probability of assignment to each potential source population across all individuals within each Latino population (see Materials and Methods for details).

**Table S1:** Summary of Latino populations and assembled reference panels

**Table S2:** Correlation p-values of male vs. female ancestry

**Table S3**. $F_{ST}$ divergences between estimated populations for K=8 using ADMIXTURE

**Table S4**. $F_{ST}$ divergences between estimated populations for K=20 using ADMIXTURE

**Text S1.** Methodology of the Ancestry-Specific PCA (ASPCA) implementation

**Text S2.** Measuring pairwise IBD between European and Latino populations

**Tables:**

**Table 1**

*Models of Migration into the Caribbean after the advent of admixture*

| **Admixed Population** | **Migration models[1]** | | | | | |
| --- | --- | --- | --- | --- | --- | --- |
| | EUR,NAT +AFR | | EUR,NAT +AFR +EUR | | EUR,NAT +AFR +AFR | |
| | Log Likelihood | Time (G)[2] | Log Likelihood | Time (G)[2] | Log Likelihood | Time (G)[2] |
| COL | -255.33 | 13 | **-246.80** | **14** | -247.68 | 13 |
| HON | -153.24 | 13 | **-139.22** | **14** | -156.03 | 13 |
| CUB | -506.43 | 19 | -497.62 | 21 | **-326.12** | **17** |
| DOM | -189.39 | 17 | -189.33 | 17 | **-170.14** | **16** |
| HAI | -122.73 | 11 | -121.91 | 12 | **-119.10** | **16** |
| PUR | -222.82 | 17 | -204.23 | 17 | **-176.17** | **16** |

[1]Three migration models were tested for each admixed population: a simple model of single pulses of migrants from each source population, beginning with Europeans and Native Americans at $T_1$ followed by African migrants at $T_2$ (EUR,NAT +AFR); the simple model followed by an additional pulse of European migrants (EUR,NAT +AFR +EUR); the simple model followed by an additional pulse of African migrants (EUR,NAT +AFR +AFR). Log likelihoods given either model were compared and we present the model with the best Bayesian Information Criterion (log likelihood values in bold).
[2]The maximum likelihood estimate of time since admixture initially. We assume prior migration between the populations was zero. Time since migration began is indicated in generations.

Figure 1.

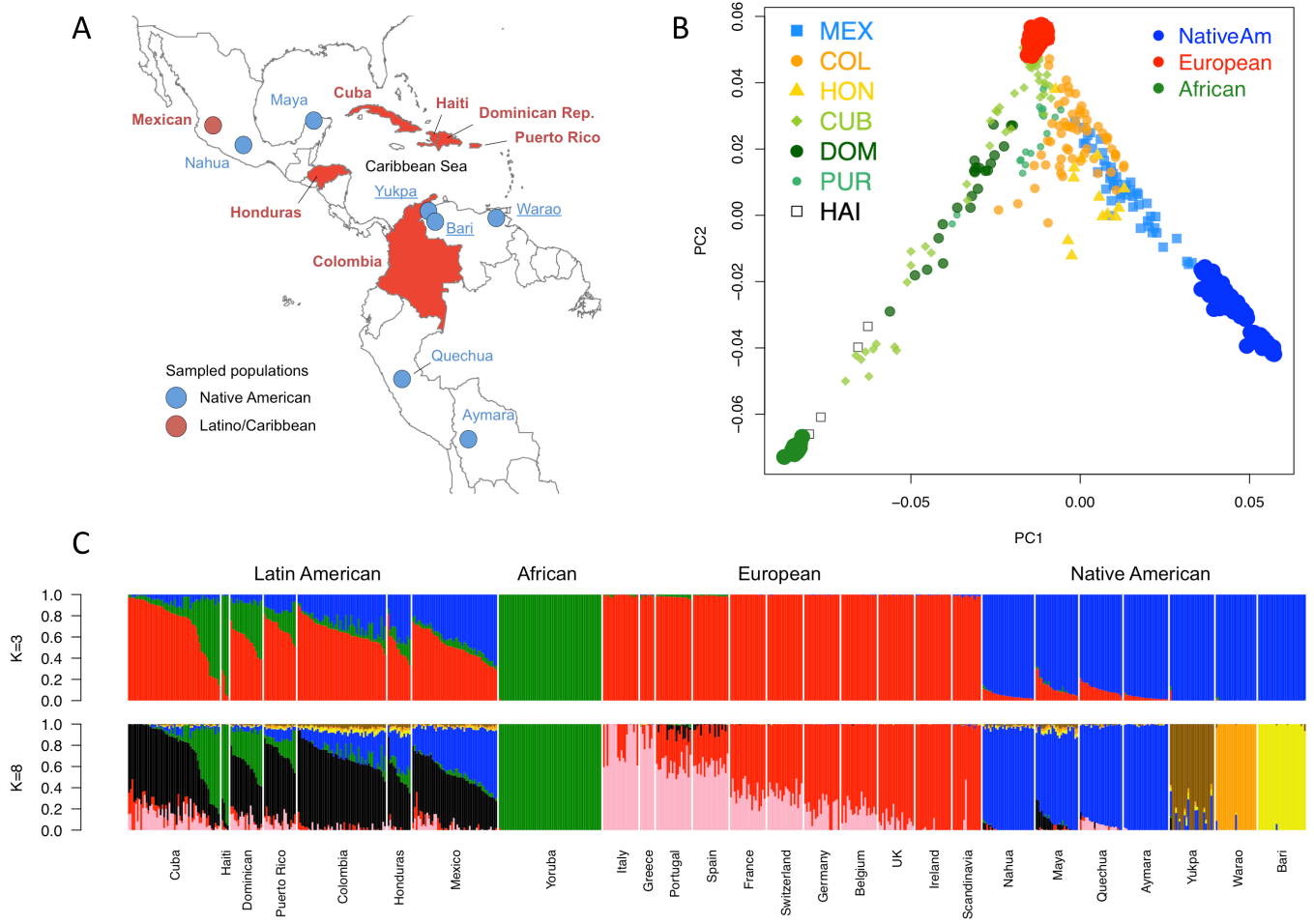

Figure 2.

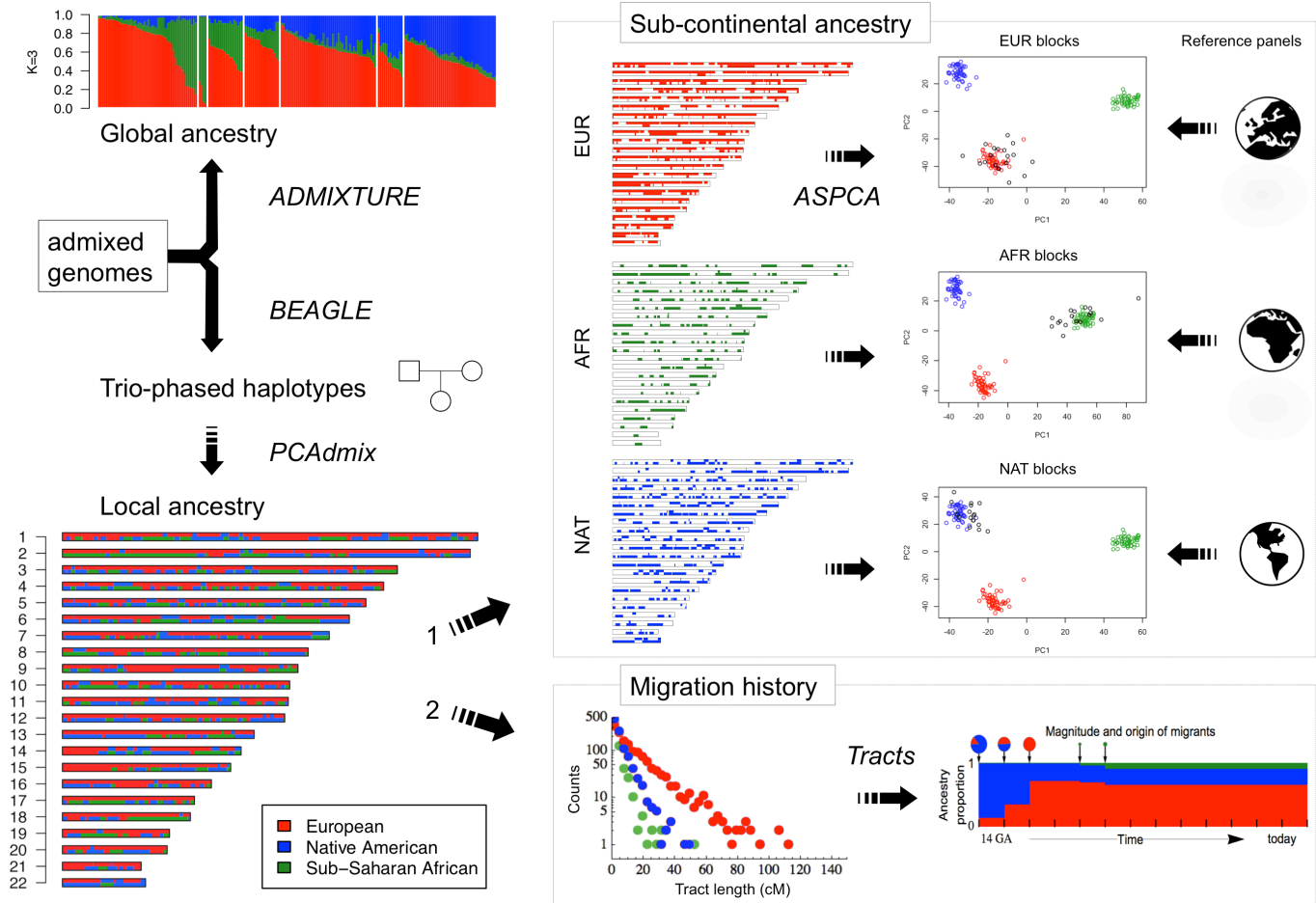

Figure 3.

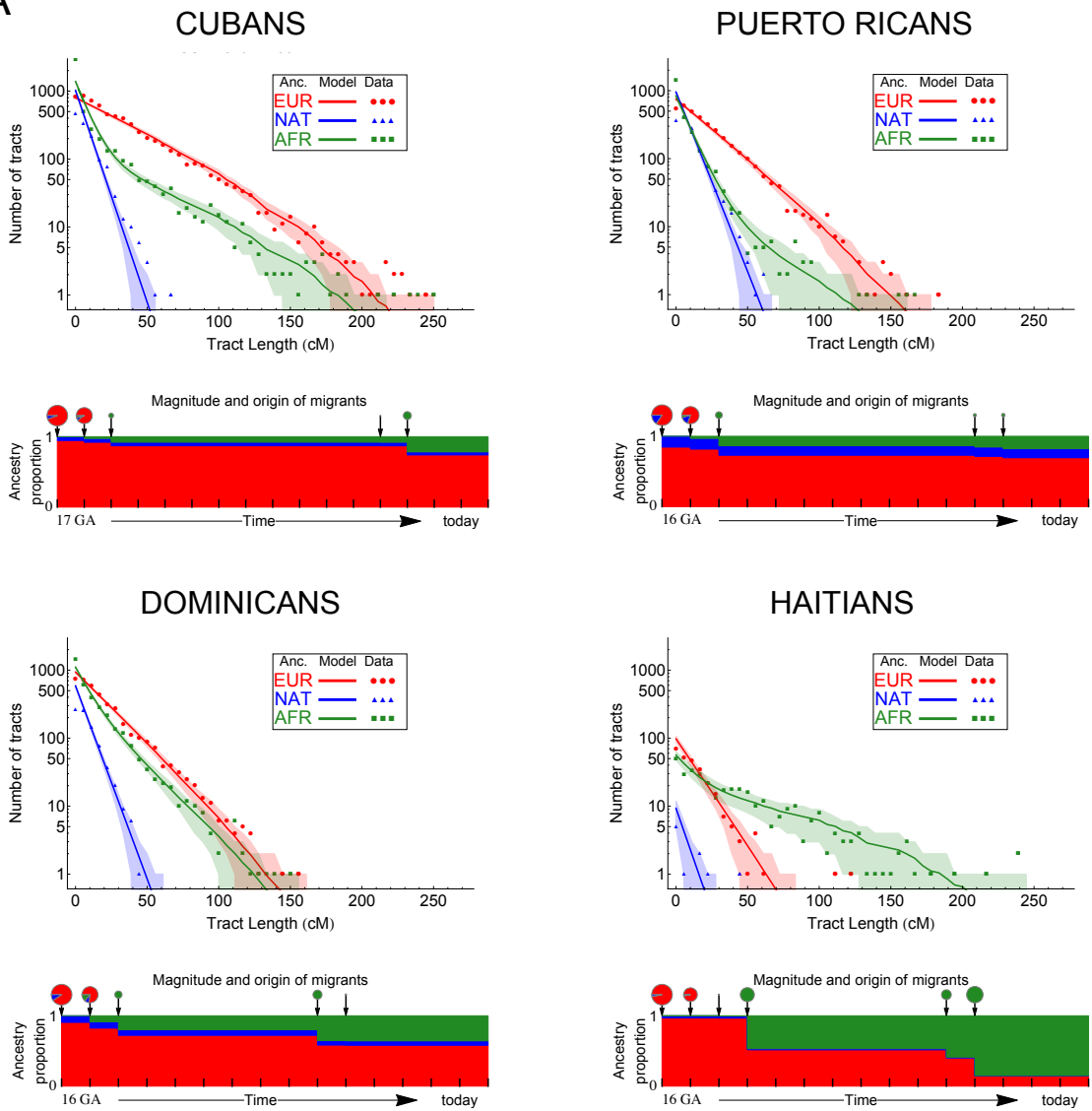

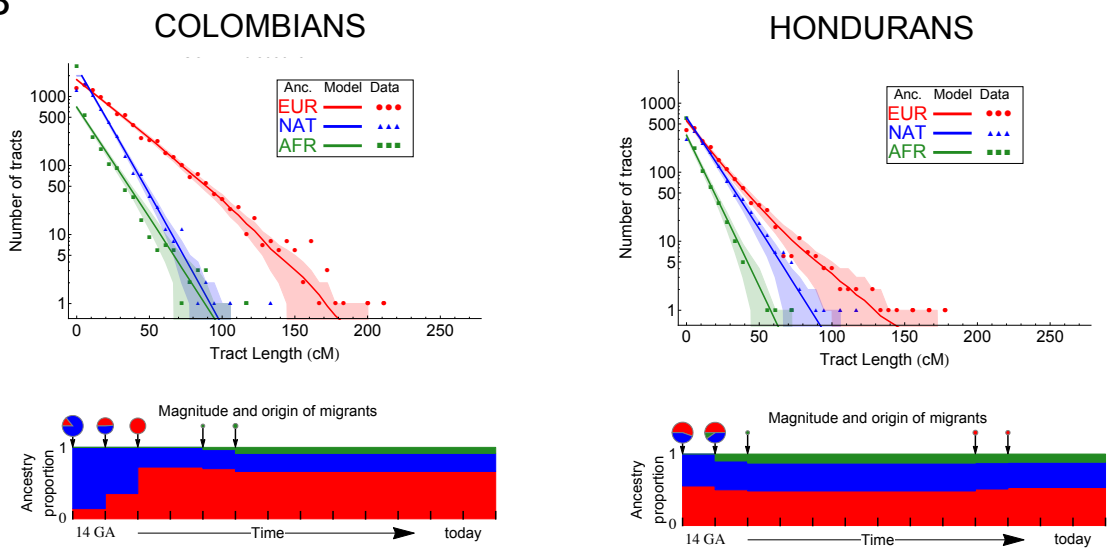

Figure 4.

Figure 5.

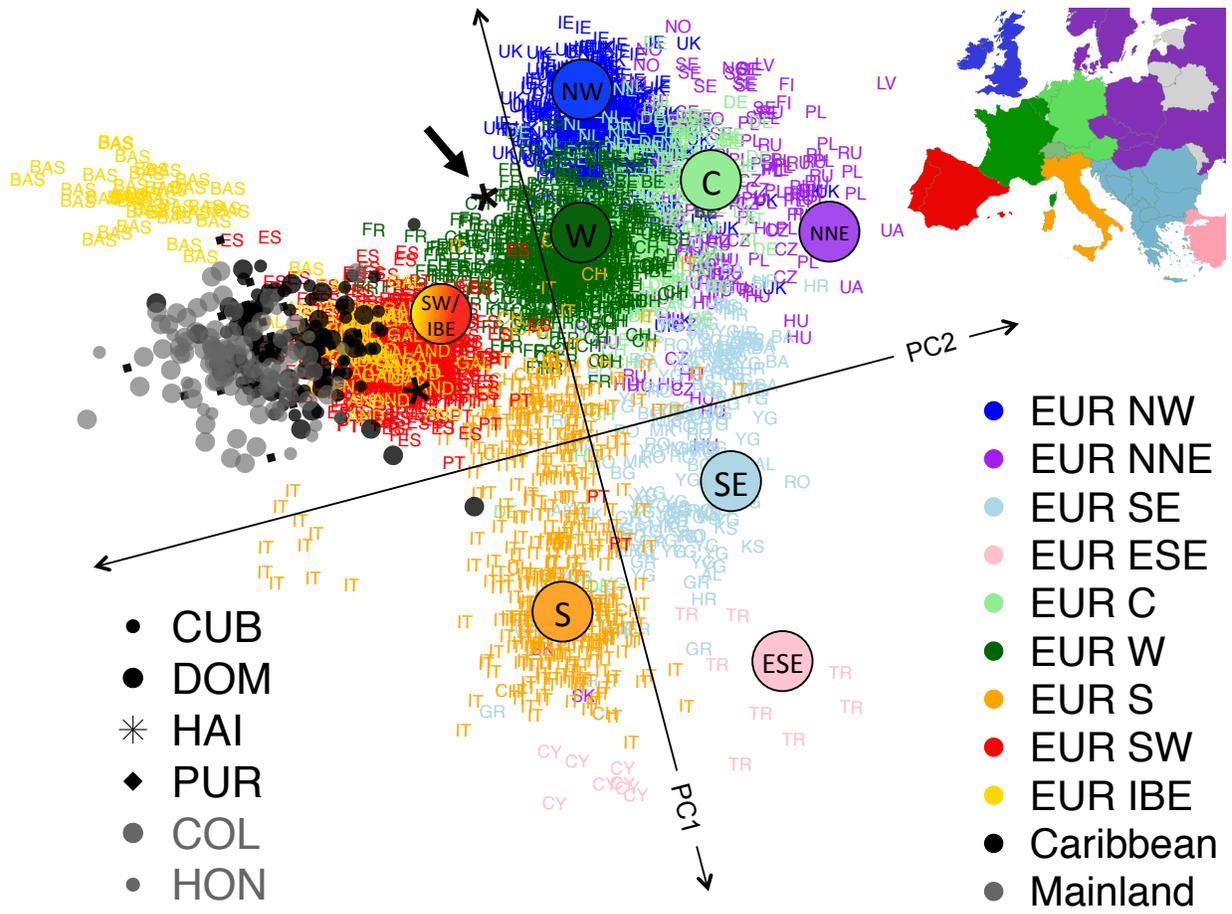

Figure 6.

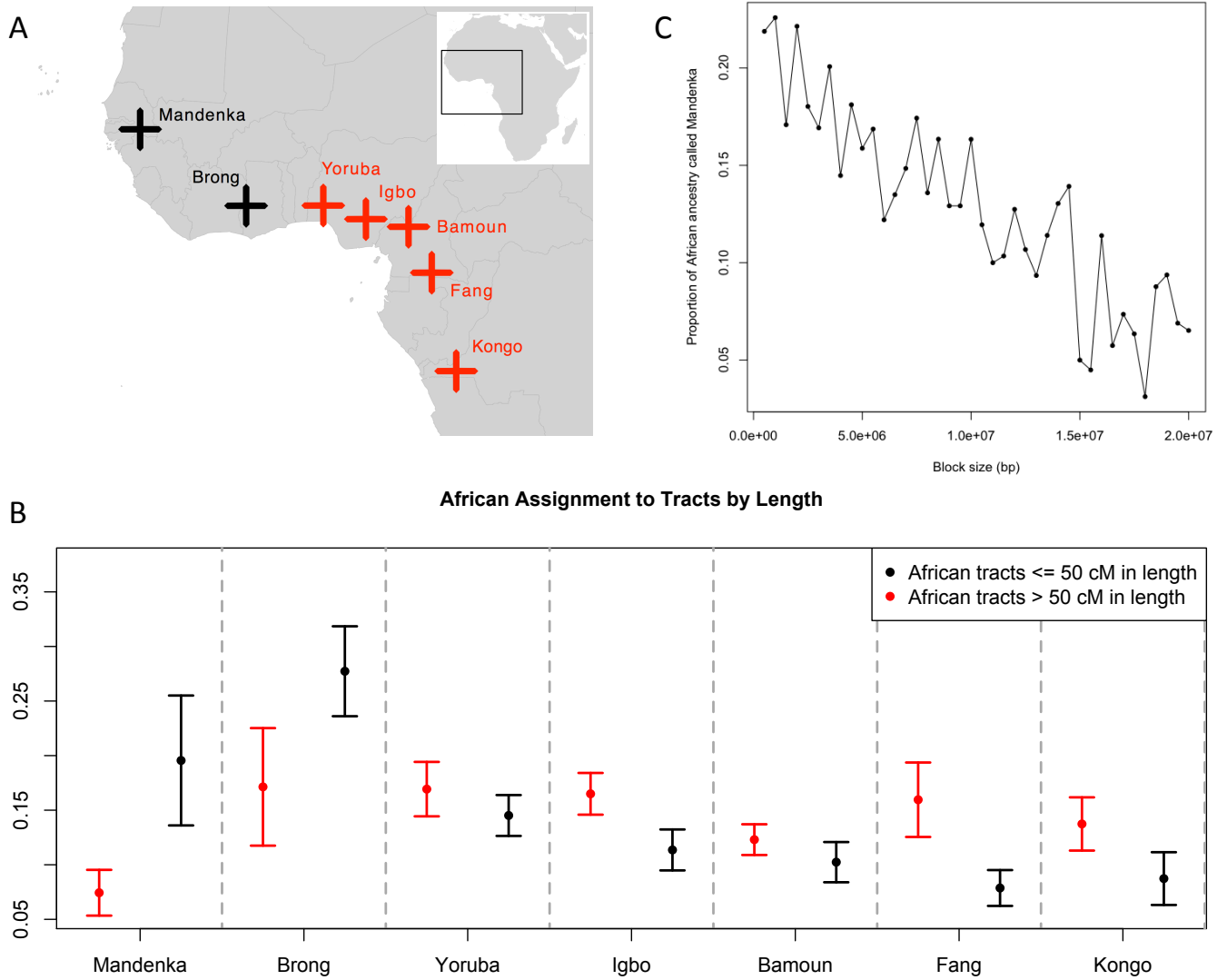

SUPPORTING INFORMATION

# Reconstructing the Population Genetic History of the Caribbean

Andrés Moreno-Estrada[1], Simon Gravel[1,2], Fouad Zakharia[1], Jacob L. McCauley[3], Jake K. Byrnes[1,4], Christopher R. Gignoux[5], Patricia A. Ortiz-Tello[1], Ricardo J. Martínez[3], Dale J. Hedges[3], Richard W. Morris[3], Celeste Eng[5], Karla Sandoval[1], Suehelay Acevedo-Acevedo[6], Juan Carlos Martínez-Cruzado[6], Paul J. Norman[7], Zulay Layrisse[8], Peter Parham[7], Esteban González Burchard[5], Michael L. Cuccaro[3], Eden R. Martin[3*], Carlos D. Bustamante[1*]

[1]Department of Genetics, Stanford University School of Medicine, Stanford, CA, USA
[2]Department of Human Genetics and Genome Quebec Innovation Centre, McGill University, Montreal, QC, Canada
[3]Center for Genetic Epidemiology and Statistical Genetics, John P. Hussman Institute for Human Genomics, University of Miami Miller School of Medicine, Miami, FL, USA
[4]Ancestry.com, San Francisco, CA, USA
[5]Department of Bioengineering and Therapeutic Sciences, University of California San Francisco, CA, USA
[6]Department of Biology, University of Puerto Rico at Mayaguez, Puerto Rico
[7]Department of Structural Biology, Stanford University School of Medicine, Stanford, CA, USA
[8]Center of Experimental Medicine "Miguel Layrisse", IVIC, Caracas, Venezuela
*Shared senior authorship and co-corresponding authors

Figure S1.

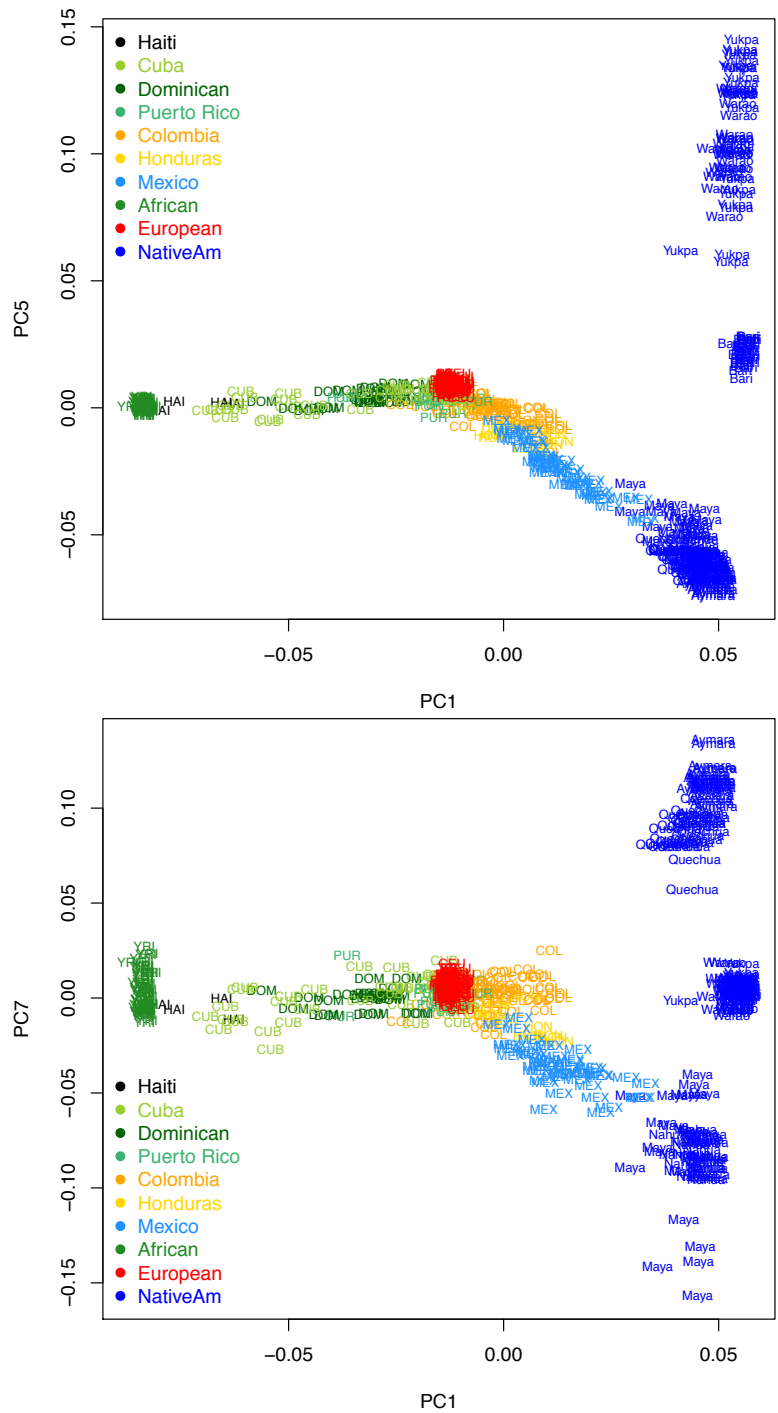

Figure S2.

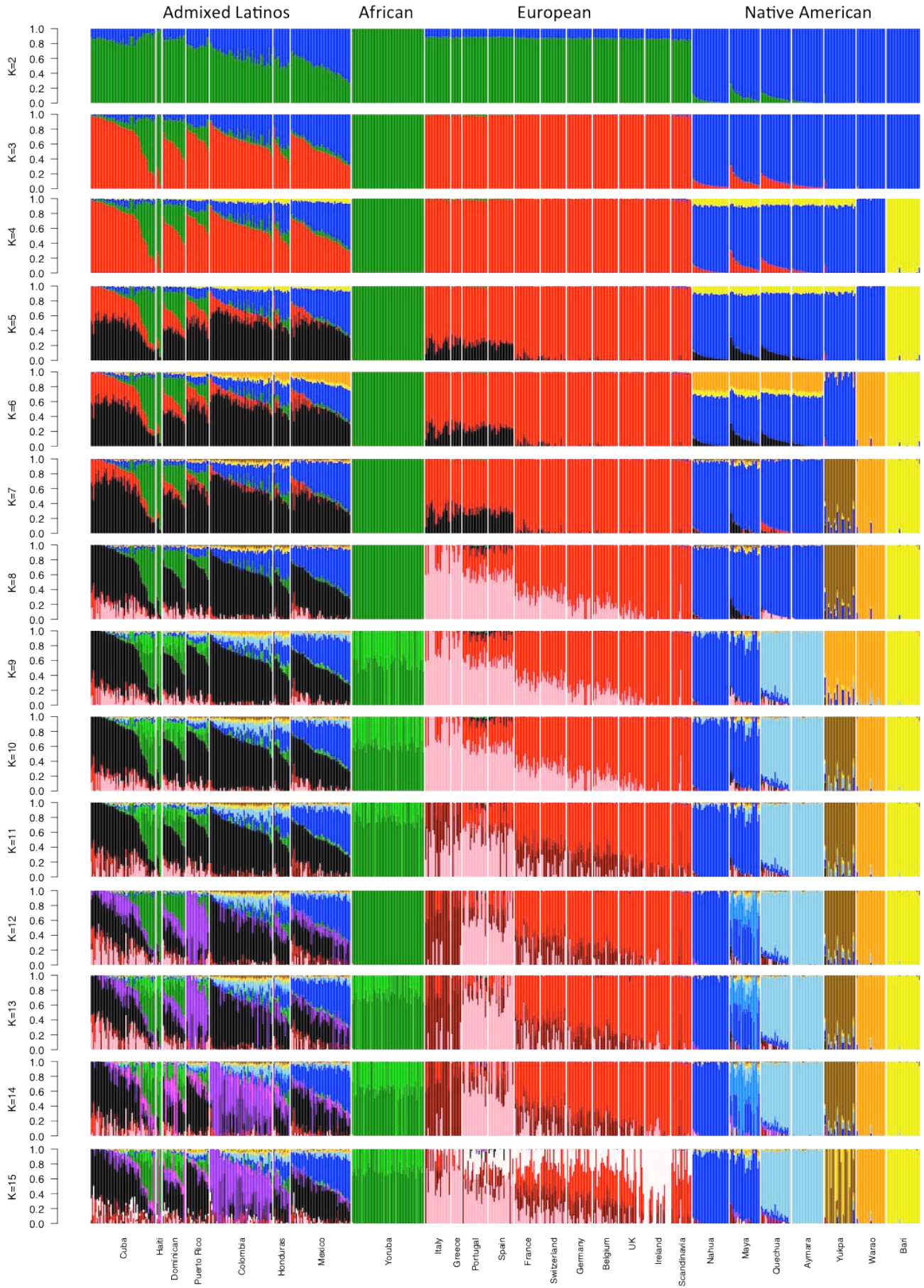

Figure S3.

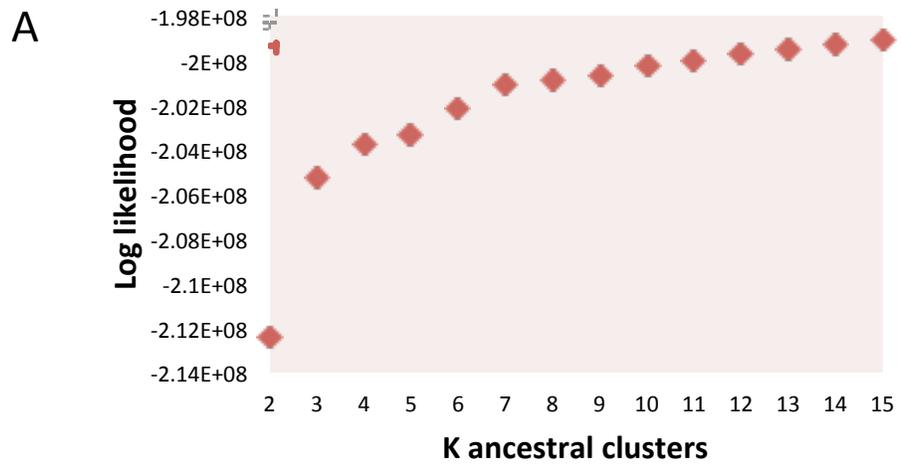

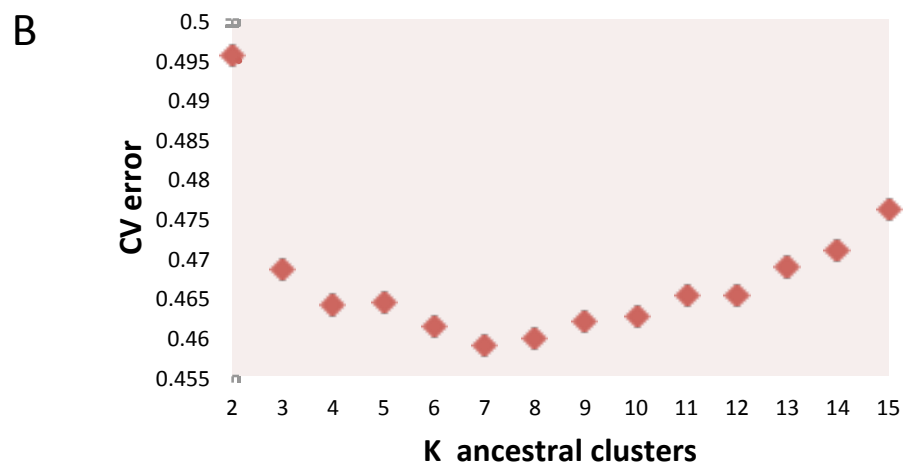

Figure S4.

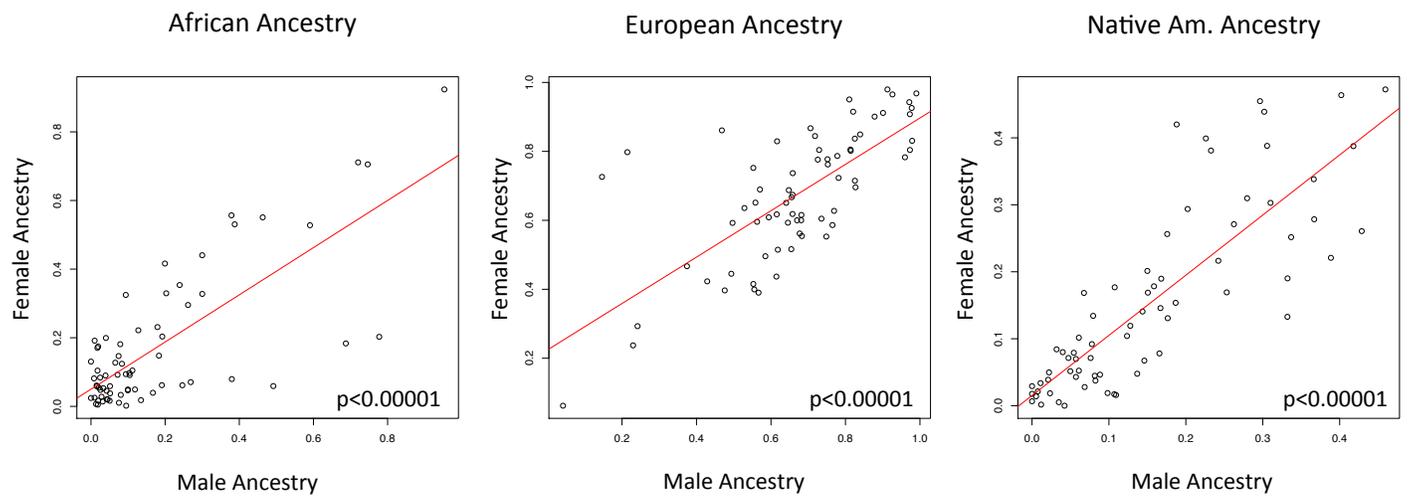

Figure S5.

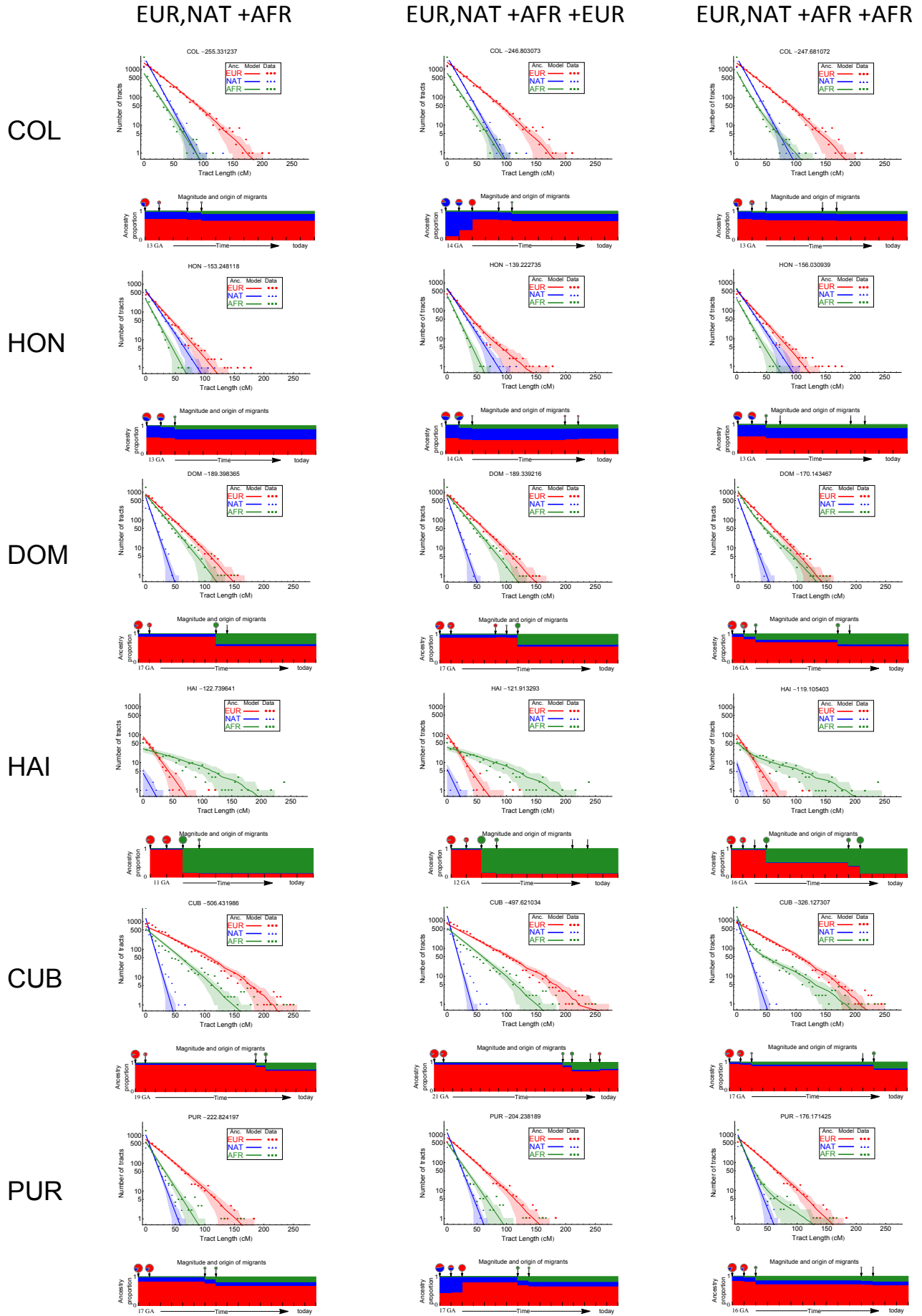

Figure S6.

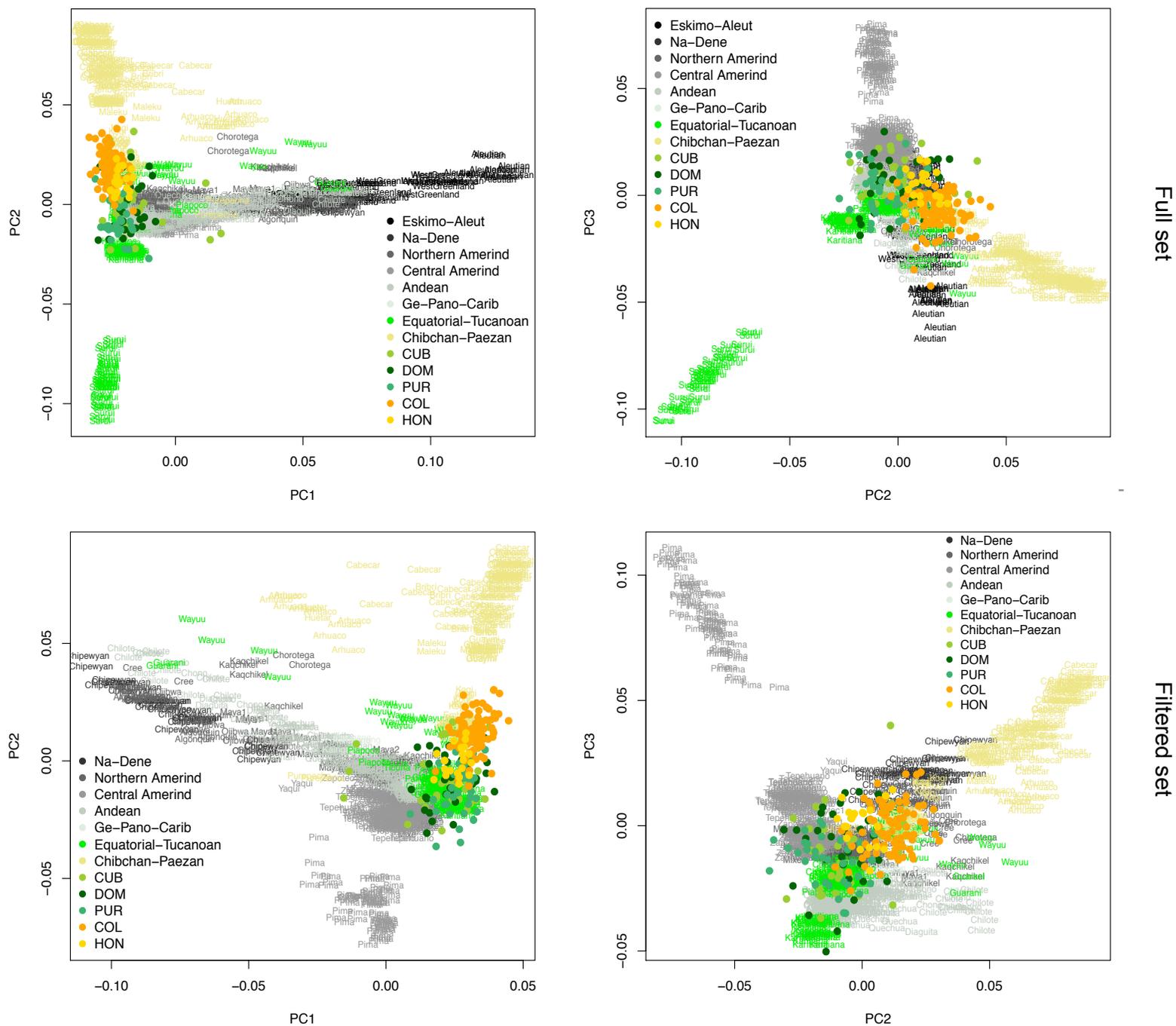

Figure S7.

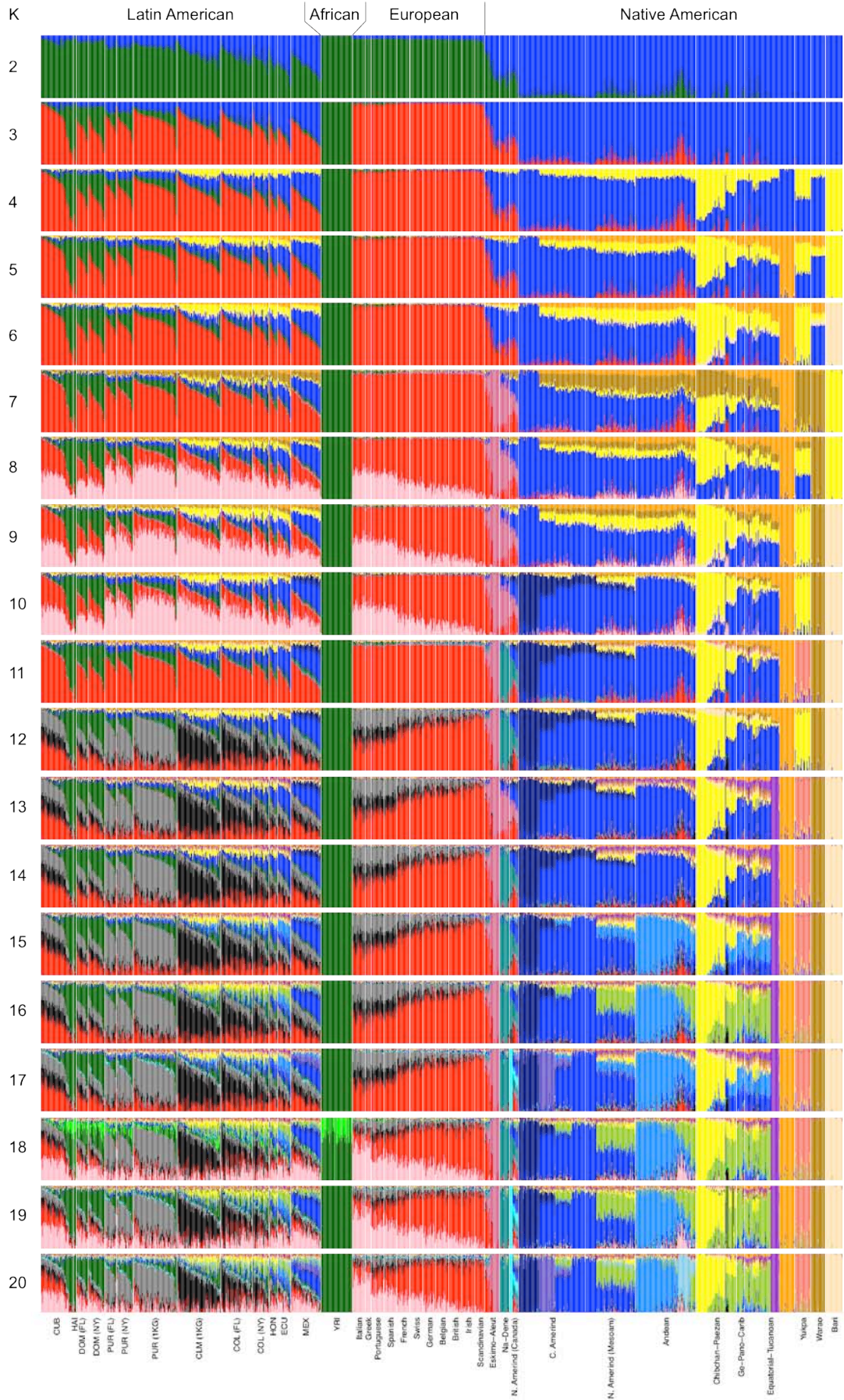

Figure S8.

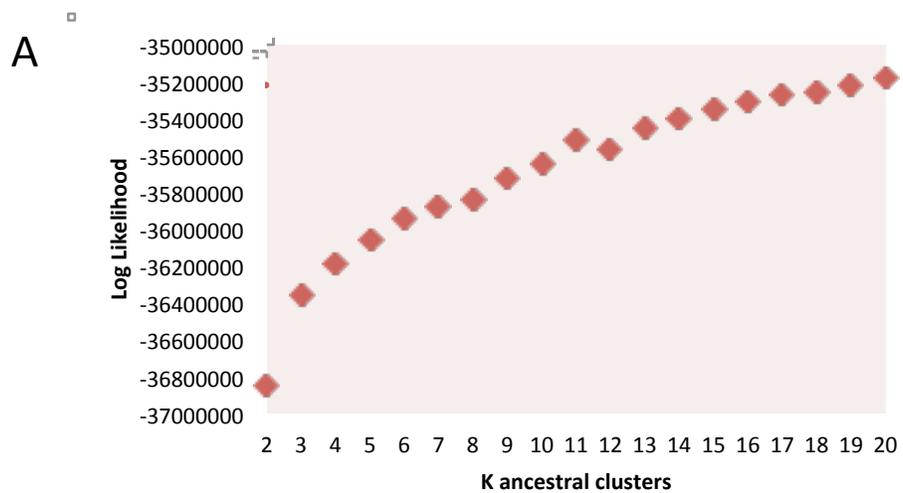

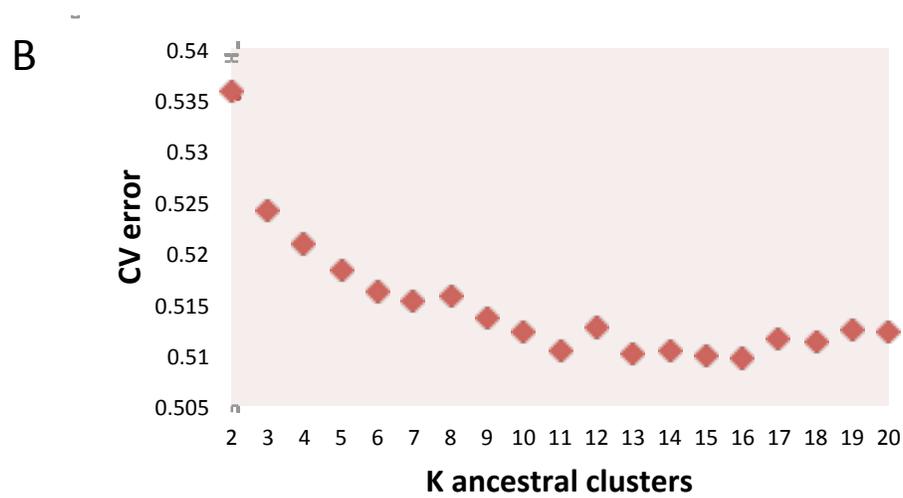

Figure S9.

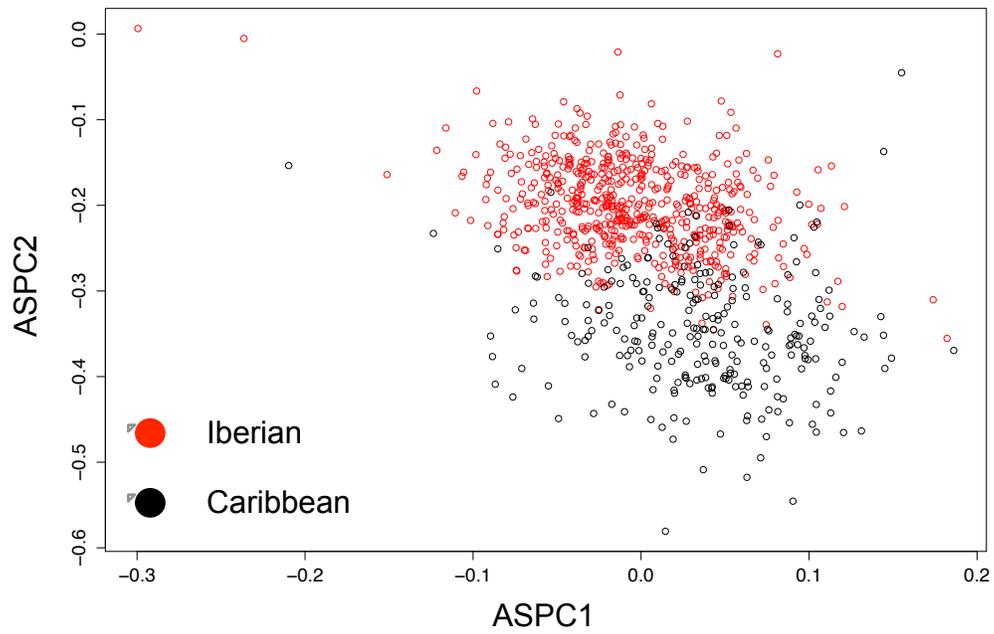

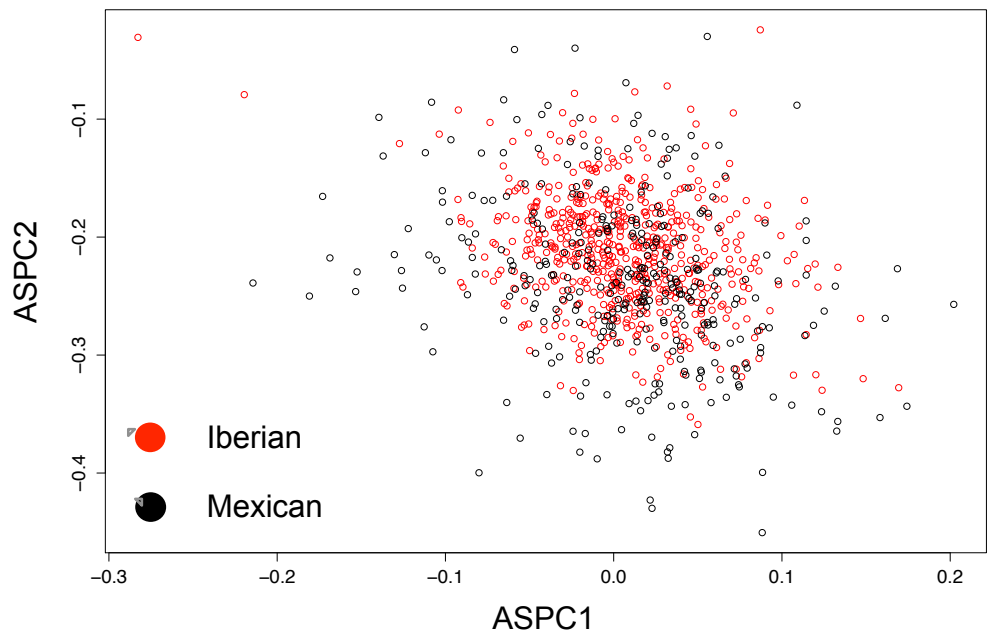

Figure S10.

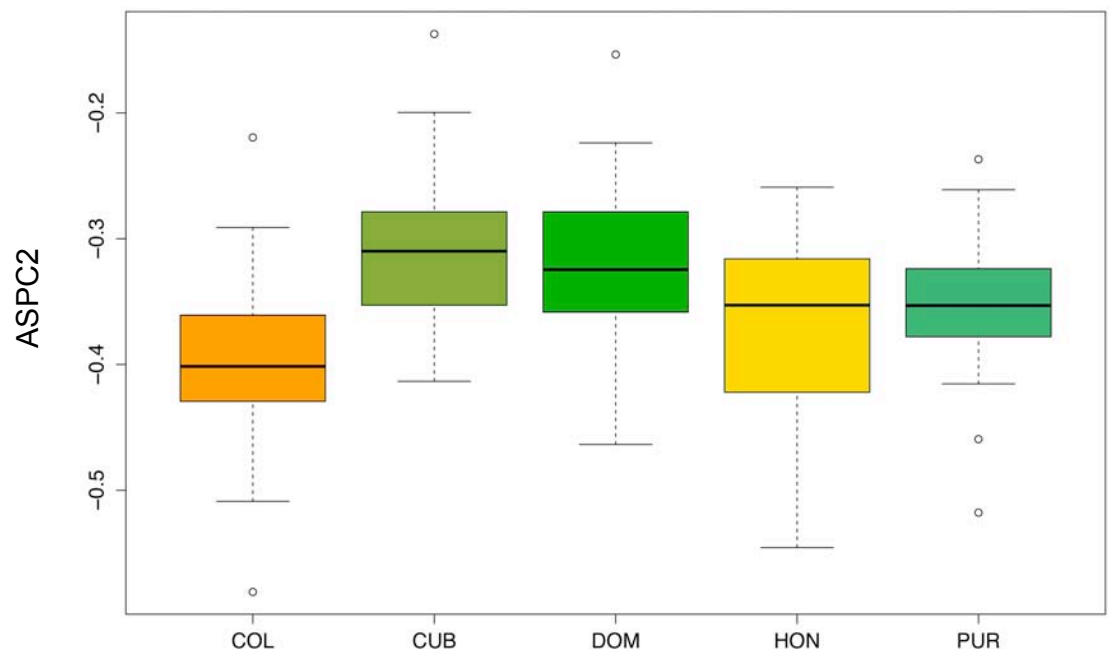

Figure S11.

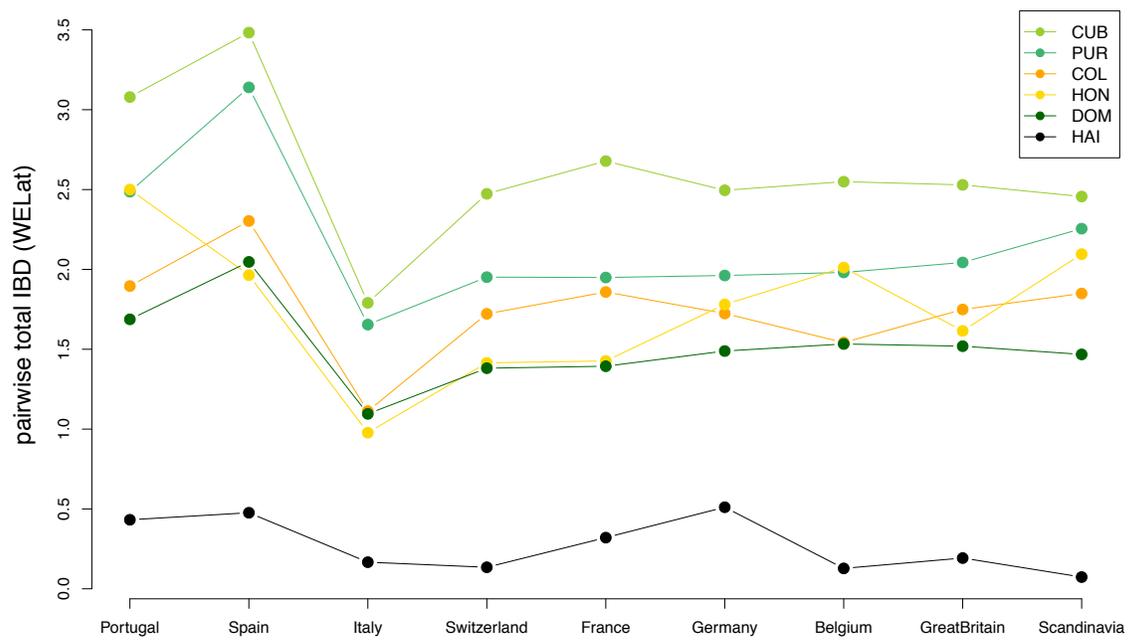

Figure S12.

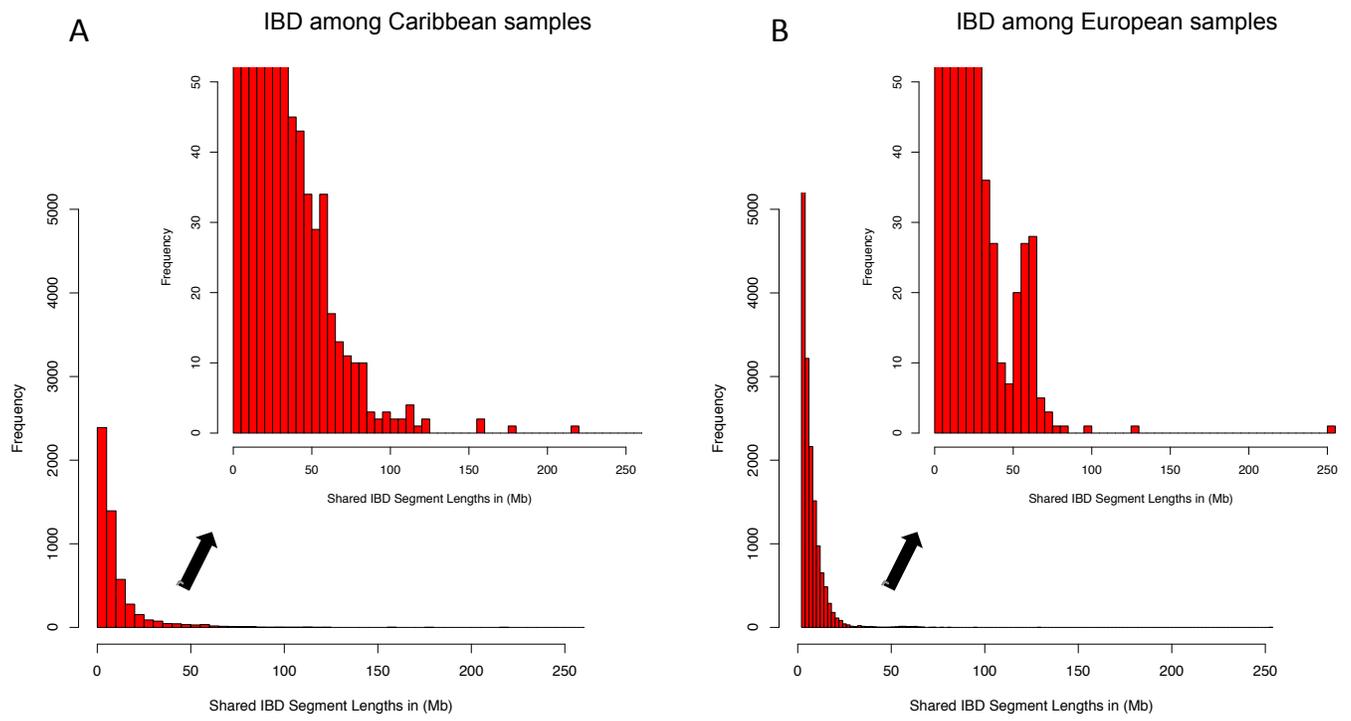

Figure S13.

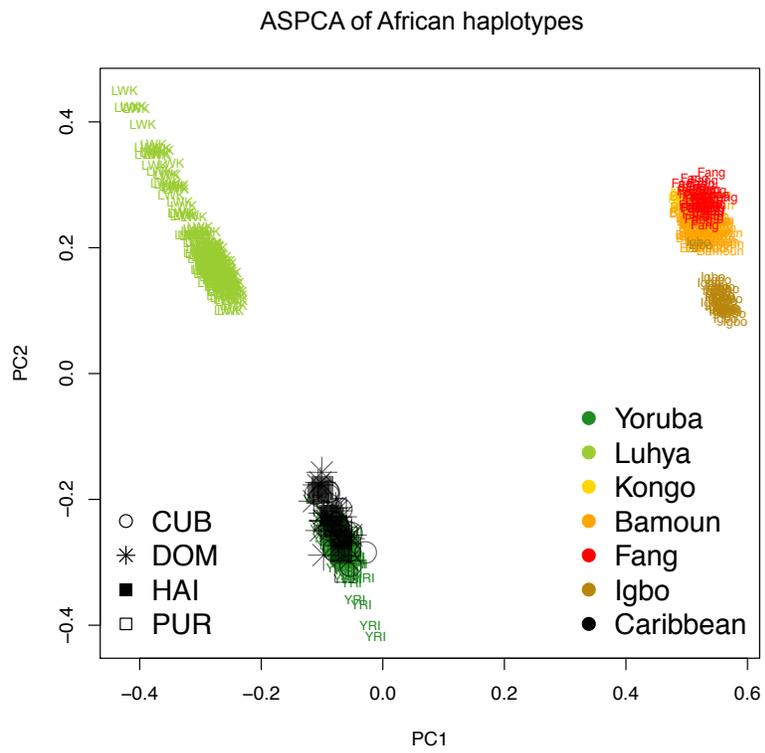

Figure S14.

Size-based ASPCA analysis of African haplotypes

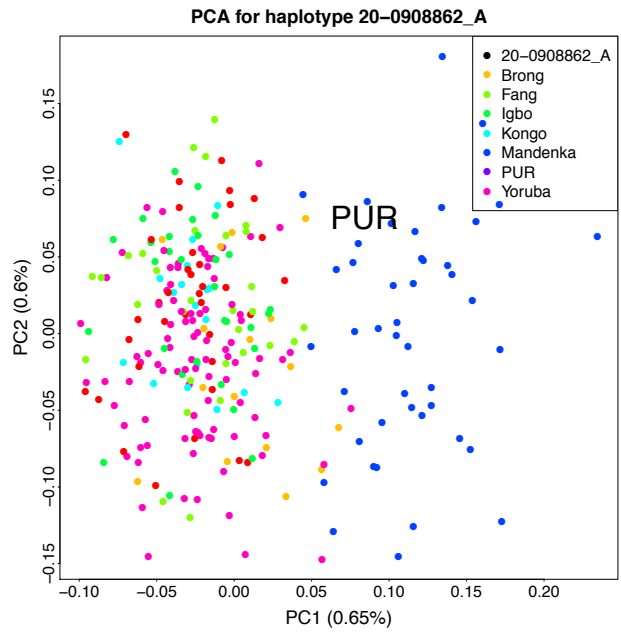
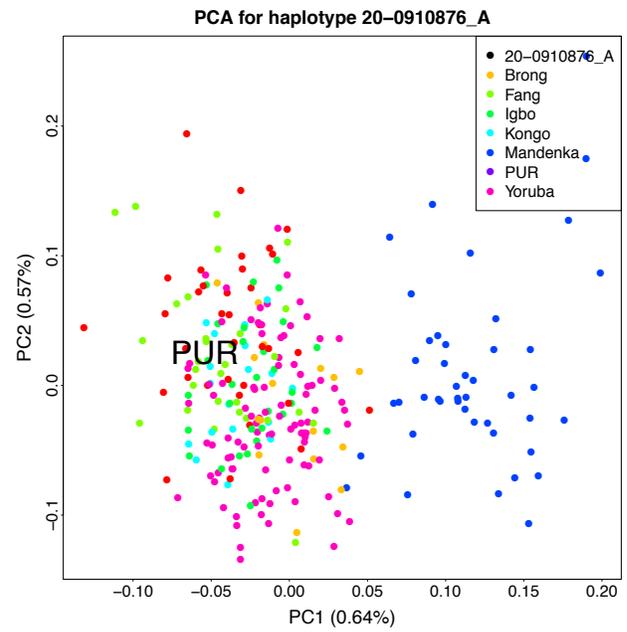

SHORT TRACTS (<50 cM)    LONG TRACTS (<50 cM)

Figure S15.

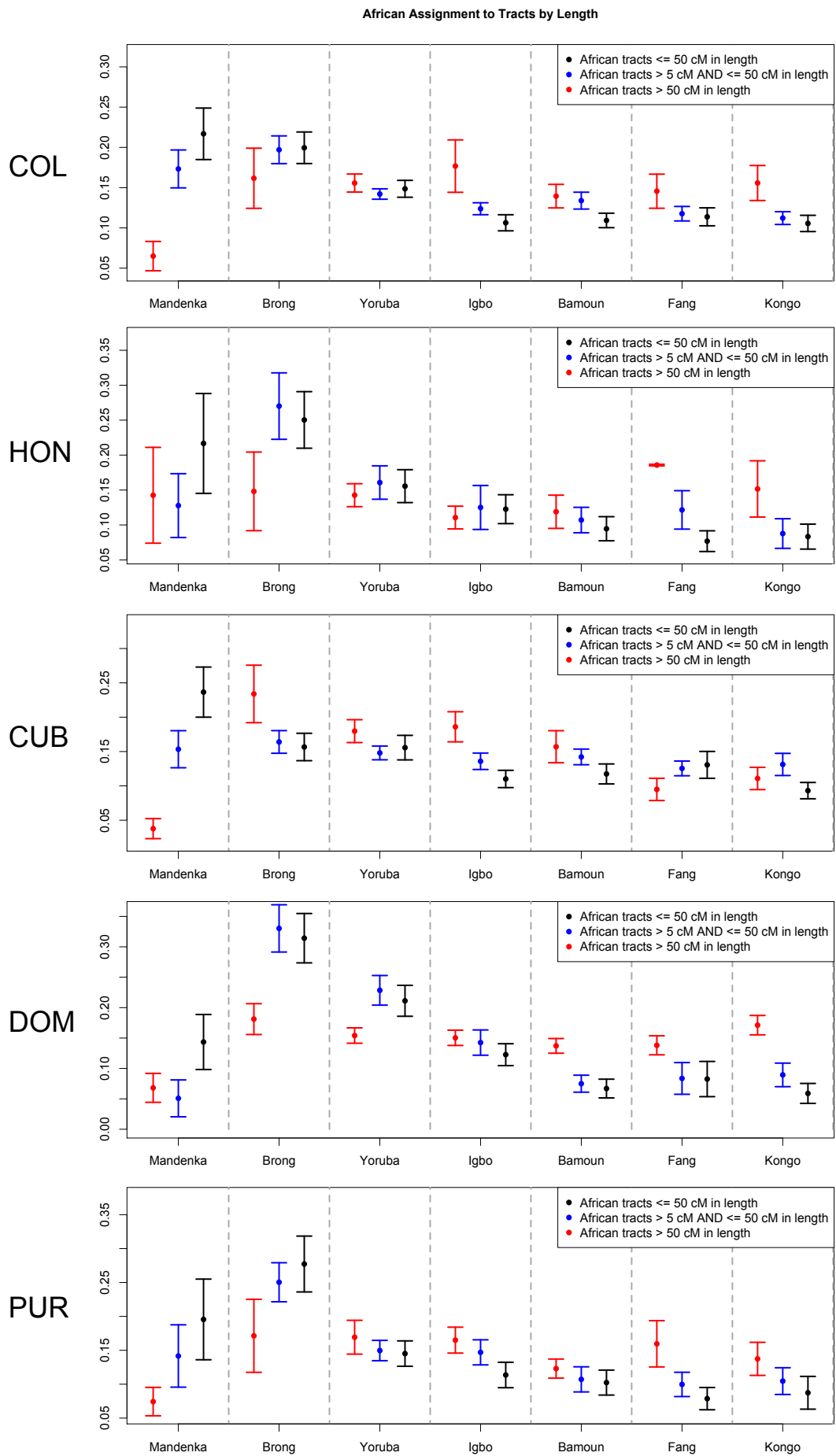

**Table S1**.

*Summary of Latino populations and assembled reference panels*

| | Population | Pop ID | Sample size[1] | Subset in low density ADMIXTURE[2] | Subset in high density ADMIXTURE[3] | Region/country | Array platform | Study/source |
|---|---|---|---|---|---|---|---|---|
| ADMIXED LATINO | Cuban in South Florida | CUB | 80 | 52 | 52 | - | affymetrix 6.0 | *Present study* |
| | Colombian in South Florida | COL (FL) | 85 | 50 | 50 | - | affymetrix 6.0 | *Present study* |
| | Dominican in South Florida | DOM (FL) | 34 | 18 | 18 | - | affymetrix 6.0 | *Present study* |
| | Puerto Rican in South Florida | PUR (FL) | 27 | 18 | 18 | - | affymetrix 6.0 | *Present study* |
| | Honduran in South Florida | HON | 19 | 13 | 13 | - | affymetrix 6.0 | *Present study* |
| | Haitian in South Florida | HAI | 6 | 4 | 4 | - | affymetrix 6.0 | *Present study* |
| | TOTAL SOUTH FLORIDA | | 251 | 155 | 155 | | | |
| | Puerto Rican in New York | PUR (NY) | 27 | 26 | - | - | illumina 650K | Bryc et al. 2010b |
| | Dominican in New York | DOM (NY) | 27 | 26 | - | - | illumina 650K | Bryc et al. 2010b |
| | Colombian in New York | COL (NY) | 26 | 26 | - | - | illumina 650K | Bryc et al. 2010b |
| | Ecuadorian in New York | ECU | 20 | 20 | - | - | illumina 650K | Bryc et al. 2010b |
| | Colombian in Medellin | CLM (1KG) | 70 | 70 | - | - | illumina Omni 2.5 | 1000 Genomes |
| | Puerto Rican in Puerto Rico | PUR (1KG) | 70 | 70 | - | - | illumina Omni 2.5 | 1000 Genomes |
| | Mexican in Los Angeles | MXL | 80 | 48 | 48 | - | affymetrix 6.0 | HapMap3 |
| | TOTAL ADMIXED LATINOS | | 571 | 441 | 203 | | | |
| NATIVE AMERICAN | Nahua | - | 30 | - | 29 | Mesoamerica | affymetrix 500K | Mao et al. 2007 |
| | Maya | - | 25 | - | 24 | Mesoamerica | affymetrix 500K | Mao et al. 2007 |
| | Quechua | - | 25 | - | 24 | Andes | affymetrix 500K | Mao et al. 2007 |
| | Aymara | - | 25 | - | 25 | Andes | affymetrix 500K | Mao et al. 2007 |
| | Bari | - | 29 | 27 | 27 | Venezuela | affymetrix 6.0 | *Present study* |
| | Yukpa | - | 25 | 25 | 25 | Venezuela | affymetrix 6.0 | *Present study* |
| | Warao | - | 25 | 23 | 23 | Venezuela | affymetrix 6.0 | *Present study* |
| | Eskimo-Aleut | - | 23 | 23 | - | North America | illumina 650K | Reich et al. 2012 |
| | Na-Dene | - | 15 | 15 | - | North America | illumina 650K | Reich et al. 2012 |
| | Northern Amerind | - | 93 | 93 | - | North America | illumina 650K | Reich et al. 2012 |
| | Central Amerind | - | 108 | 108 | - | North America | illumina 650K | Reich et al. 2012 |
| | Chibchan-Paezan | - | 65 | 65 | - | Central America | illumina 650K | Reich et al. 2012 |
| | Andean | - | 97 | 97 | - | South America | illumina 650K | Reich et al. 2012 |
| | Ge-Pano-Carib | - | 12 | 12 | - | South America | illumina 650K | Reich et al. 2012 |
| | Equatorial-Tucanoan | - | 80 | 80 | - | South America | illumina 650K | Reich et al. 2012 |
| | TOTAL NATIVE AMERICAN[4] | | 677 | 568 | 177 | | | |
| EUROPEAN | European (North West) | EUR NW | 266 | 40 | 40 | - | affymetrix 500K | POPRES[6] |
| | European (North/Norh East) | EUR NNE | 76 | 15 | 15 | - | affymetrix 500K | POPRES |
| | European (South East) | EUR SE | 96 | 8 | 8 | - | affymetrix 500K | POPRES |
| | European (East/South East) | EUR ESE | 8 | - | - | - | affymetrix 500K | POPRES |
| | European (Central) | EUR C | 186 | 20 | 20 | - | affymetrix 500K | POPRES |
| | European (West) | EUR W | 259 | 60 | 60 | - | affymetrix 500K | POPRES |
| | European (South) | EUR S | 232 | 20 | 20 | - | affymetrix 500K | POPRES |
| | European (South West) | EUR SW | 264 | 40 | 40 | - | affymetrix 500K | POPRES |
| | TOTAL POPRES[5] | | 1387 | 203 | 203 | | | |
| | Iberian in Andalusia | IBE AND | 17 | - | - | Spain | affymetrix 6.0 | Rodriguez et al. (in revision) |
| | Iberian in Galicia | IBE GAL | 17 | - | - | Spain | affymetrix 6.0 | Rodriguez et al. (in revision) |
| | Iberian in the Basque Country | IBE BAS | 20 | - | - | Spain | affymetrix 6.0 | Henn et al. 2012 |
| | TOTAL EUROPEAN | | 1441 | 203 | 203 | | | |
| AFRICAN | Yoruba | YRI | 167 | 50 | 58 | West Africa | affymetrix 6.0 | HapMap3 |
| | Luhya | LWK | 90 | - | - | East Africa | affymetrix 6.0 | HapMap3 |
| | Bamoun | - | 20 | - | - | West Africa | affymetrix 500K | Bryc et al. 2010a |
| | Fang | - | 18 | - | - | West Africa | affymetrix 500K | Bryc et al. 2010a |
| | Igbo | - | 17 | - | - | West Africa | affymetrix 500K | Bryc et al. 2010a |
| | Kongo | - | 9 | - | - | West Africa | affymetrix 500K | Bryc et al. 2010a |
| | Brong | - | 8 | - | - | West Africa | affymetrix 500K | Bryc et al. 2010a |
| | Mandenka | - | 24 | - | - | West Africa | illumina 650K | HGDP |
| | TOTAL AFRICAN | | 353 | 50 | 58 | | | |
| | TOTAL | | 3042 | 1262 | 641 | | | |

[1]Total of samples included in the study. For Native American, European, and African populations, this is the maximum number of samples used to construct the reference panels (e.g., for ASPCA analyses).

[2]The low-density dataset consists of 30,860 SNPs from a representative subset of most populations with Affymetrix SNP array data merged with populations with Illumina SNP array data. Numbers vary from the initial sample size either due to QC filtering (e.g., trios offspring, cryptic related individuals, and PCA outliers) or to broadly equalize sample sizes across populations (e.g., within POPRES collection).

[3]The high-density dataset consists of 389,225 SNPs from a representative subset of most populations with Affymetrix SNP array data. Only a subset of YRI samples were included in order to focus on European and Native American sub-continental structure (other analyses, such as ASPCA, were performed focusing on sub-continental African ancestry including all West African populations).

[4]The subset of 493 Native American samples from Reich et al (2012) represent a total of 52 populations. However, for summary purposes, these are shown grouped by linguistic families as in the original publication.

[5]We restricted to 1,387 POPRES European samples with four grandparents from the same country as in Novembre et al. (2008) to ensure replication of the PCA map of Europe. Geographic groups are as in Auton et al (2009). Full details of studied populations by country are available in Novembre et al. (2008).

[6]The collections and methods for the Population Reference Sample (POPRES) are described by Nelson et al. (2008). The datasets used for the analyses described in this manuscript were obtained from dbGaP at http://www.ncbi.nlm.nih.gov/projects/gap/cgibin/study.cgi?study_id=phs000145.v1.p1 through dbGaP accession number phs000145.v1.p1.

**Table S2**

*Correlation p-values of male vs. female ancestry[1]*

| Population | AFR       | EUR       | NAT       |
|------------|-----------|-----------|-----------|
| COL        | 0.00022   | 9.00E-05  | 1.00E-05  |
| CUB        | 0.00023   | 0.00046   | < 0.00001 |
| DOM        | 0.03748   | 0.03355   | 0.01294   |
| HAI        | 0.49834   | 0.49923   | 0.49835   |
| HON        | 0.01407   | < 0.00001 | < 0.00001 |
| PUR        | 0.04182   | 0.01473   | 0.01058   |
| ALL        | < 0.00001 | < 0.00001 | < 0.00001 |

[1]Significance was assessed by comparing correlation of ancestry assignments among parent pairs to 100,000 permuted male-female pairs in each population.

**Table S3**

*$F_{ST}$ divergences between estimated populations for K=8 using ADMIXTURE*

| K=8 | Eur-North | Yukpa | Warao | Bari | Eur-South | Eur-Latino | Amerind |
|---|---|---|---|---|---|---|---|
| Yukpa | 0.274 | | | | | | |
| Warao | 0.26 | 0.232 | | | | | |
| Bari | 0.277 | 0.239 | 0.234 | | | | |
| Eur-South | 0.02 | 0.284 | 0.269 | 0.286 | | | |
| Eur-Latino | 0.015 | 0.264 | 0.25 | 0.266 | 0.021 | | |
| Amerind | 0.165 | 0.133 | 0.118 | 0.136 | 0.174 | 0.158 | |
| Yoruba | 0.184 | 0.372 | 0.358 | 0.375 | 0.176 | 0.164 | 0.27 |

**Table S4**

*F$_{ST}$ divergences between estimated populations for K=20 using ADMIXTURE*

| K=20 | Eur-North | Surui | Chipewyan | Pima | Yukpa | Mixe | Inuit | Patagonia | Eur-Latino insular | Eur-South | Karitiana | Yoruba | Bari | Eur-Latino mainland | Cabecar | Andean | Tepehuan | Ticuna | Algonquin |
|---|---|---|---|---|---|---|---|---|---|---|---|---|---|---|---|---|---|---|---|
| Surui | 0.296 | | | | | | | | | | | | | | | | | | |
| Chipewyan | 0.159 | 0.237 | | | | | | | | | | | | | | | | | |
| Pima | 0.227 | 0.214 | 0.162 | | | | | | | | | | | | | | | | |
| Yukpa | 0.29 | 0.259 | 0.229 | 0.206 | | | | | | | | | | | | | | | |
| Mixe | 0.19 | 0.157 | 0.121 | 0.089 | 0.151 | | | | | | | | | | | | | | |
| Inuit | 0.191 | 0.259 | 0.153 | 0.185 | 0.256 | 0.144 | | | | | | | | | | | | | |
| Patagonia | 0.175 | 0.178 | 0.133 | 0.121 | 0.17 | 0.071 | 0.157 | | | | | | | | | | | | |
| Eur-Latino insular | 0.037 | 0.291 | 0.165 | 0.225 | 0.287 | 0.188 | 0.193 | 0.176 | | | | | | | | | | | |
| Eur-South | 0.039 | 0.312 | 0.175 | 0.24 | 0.305 | 0.205 | 0.202 | 0.189 | 0.041 | | | | | | | | | | |
| Karitiana | 0.272 | 0.216 | 0.213 | 0.185 | 0.232 | 0.132 | 0.234 | 0.153 | 0.269 | 0.287 | | | | | | | | | |
| Yoruba | 0.185 | 0.384 | 0.26 | 0.313 | 0.376 | 0.277 | 0.278 | 0.267 | 0.157 | 0.177 | 0.362 | | | | | | | | |
| Bari | 0.291 | 0.259 | 0.229 | 0.204 | 0.239 | 0.15 | 0.251 | 0.169 | 0.288 | 0.304 | 0.231 | 0.377 | | | | | | | |
| Eur-Latino mainland | 0.035 | 0.289 | 0.161 | 0.221 | 0.279 | 0.184 | 0.19 | 0.171 | 0.045 | 0.04 | 0.264 | 0.174 | 0.278 | | | | | | |
| Cabecar | 0.259 | 0.224 | 0.194 | 0.166 | 0.195 | 0.112 | 0.218 | 0.134 | 0.256 | 0.271 | 0.197 | 0.344 | 0.177 | 0.244 | | | | | |
| Andean | 0.196 | 0.154 | 0.13 | 0.102 | 0.151 | 0.047 | 0.152 | 0.068 | 0.194 | 0.21 | 0.131 | 0.283 | 0.151 | 0.188 | 0.115 | | | | |
| Tepehuan | 0.192 | 0.171 | 0.13 | 0.094 | 0.166 | 0.054 | 0.154 | 0.086 | 0.191 | 0.206 | 0.146 | 0.279 | 0.165 | 0.186 | 0.128 | 0.066 | | | |
| Ticuna | 0.206 | 0.157 | 0.136 | 0.104 | 0.149 | 0.055 | 0.156 | 0.075 | 0.201 | 0.22 | 0.131 | 0.289 | 0.142 | 0.195 | 0.119 | 0.051 | 0.067 | | |
| Algonquin | 0.188 | 0.281 | 0.168 | 0.206 | 0.275 | 0.164 | 0.21 | 0.177 | 0.195 | 0.201 | 0.253 | 0.294 | 0.271 | 0.19 | 0.237 | 0.171 | 0.173 | 0.177 | |
| Warao | 0.278 | 0.241 | 0.214 | 0.193 | 0.236 | 0.135 | 0.238 | 0.156 | 0.276 | 0.295 | 0.215 | 0.363 | 0.234 | 0.27 | 0.202 | 0.136 | 0.151 | 0.134 | 0.258 |

**Text S1. Methodology of the Ancestry-Specific PCA (ASPCA) implementation**

Overview of the ASPCA method (subspace learning algorithm): The method we describe here is a close adaptation of the *subspace learning algorithm* described in [1] to haplotype data. In contrast to the standard approach, which computes all principal components, the subspace algorithm does away with the covariance matrix altogether, and computes the first d principal components, where $1 \leq d \leq n$. Specifically, given an m x n matrix of haplotypes, the algorithm seeks to obtain the decomposition $\mathbf{X} \approx \mathbf{AS}$, where **S** is a m x d matrix, and **A** is a d x n matrix containing the top *d* principal component loadings for every individual in the sample. For our purposes, we are interested in obtaining the latter to approximate PCA. In the absence of missing data, this decomposition can be obtained iteratively by gradient descent. Starting with random matrices **A** and **S**, the following update rules are alternatively applied to each matrix until convergence is achieved:

$$\mathbf{A} \leftarrow \mathbf{A} + \gamma(\mathbf{X} - \mathbf{AS})\mathbf{S}^T$$

$$\mathbf{S} \leftarrow \mathbf{S} + \gamma \mathbf{A}^T (\mathbf{X} - \mathbf{AS})$$

where γ controls the learning rate. Note that the resulting matrices are not necessarily orthogonal. However, orthogonalization can readily be performed post-hoc. For instance, one can orthogonalize A by SVD. Letting $\mathbf{A} = \mathbf{UDV}^T$, the orthogonalization is computed as:

$$\mathbf{A}^* = \mathbf{UV}^T$$

The progression of the algorithm towards convergence can be followed by tracking the change in the cost function C at every iteration, where C is defined as:

$$C = \sum_{i=1}^{n}\sum_{j=1}^{m}(x_{ij} - \sum_{k=1}^{d}a_{ik}s_{kj})^2$$

Intuitively, this is the mean square error between the data matrix **X** and its estimate **AS**. Throughout the algorithm, C is expected to converge to a local optimum in a monotonically decreasing fashion.

Focusing on a specific ancestry component (introduction of missing data): Given this framework, the above equations can be readily adapted to the presence of missing data, corresponding to regions of the genome that have been masked out to enable the study of a specific ancestral component of admixture. Specifically, instead of iterating over all possible entries of the haplotype matrix, we now only focus on those that are non-missing (i.e. those determined by the ancestry deconvolution algorithm to be derived from the desired admixture component). Thus, the cost function becomes:

$$C = \sum_{(i,j)\in O}(x_{ij} - \sum_{k=1}^{d}a_{ik}s_{kj})^2$$

where O now denotes the set of all observed values in the haplotype matrix X. Concordantly, the update equations corresponding to the gradient descent algorithm become [1]:

$$a_{ik} \leftarrow a_{ik} - \gamma \frac{\partial C}{\partial a_{ik}} = a_{ik} + \gamma \sum_{j|(i,j)\in O}\sum_{k=1}^{d}(x_{ij} - a_{ik}s_{kj})^2 s_{kj}$$

$$s_{kj} \leftarrow s_{kj} - \gamma \frac{\partial C}{\partial s_{kj}} = s_{kj} + \gamma \sum_{i|(i,j)\in O}\sum_{k=1}^{d}(x_{ij} - a_{ik}s_{kj})^2 a_{ik}$$

Implementation: Our implementation of the algorithm, which we packaged into the software *PCAmask*, follows the guidelines of the seminal paper quite closely [1]. Specifically, we adapted the standard gradient descent outlined above to include a speed-up term for faster convergence. We achieved this by multiplying the gradient by the inverse of the second order derivatives of the cost function, as described in [1]:

$$a_{ik} \leftarrow a_{ik} - \gamma \left( \frac{\partial^2 C}{\partial a_{ik}^2} \right)^{-1} \frac{\partial C}{\partial a_{ik}}$$

$$s_{kj} \leftarrow s_{kj} - \gamma \left( \frac{\partial^2 C}{\partial s_{kj}^2} \right)^{-1} \frac{\partial C}{\partial s_{kj}}$$

Finally, we followed the guidelines described in [1] to set the convergence term γ. At the beginning of the algorithm, we set γ = 1. At every iteration, γ is then updated based on the new value $C_{next}$ of the cost function. If $C_{next}$ < C, we set γ′ = 1.1γ; otherwise, the update of **A** and **S** is rejected and γ′ = γ/2. This approach ensures that smaller steps are taken as the process nears the local optimum.

ASPCA projection of ancestry-specific haplotypes: We used *PCAmask* (described above) to perform PCA on masked genomes of admixed origin exposing haploid loci of estimated African, European, or Native American ancestry, separately. We restricted to haploid genomes with more than 25% of European or African ancestry to be considered in the analysis. Due to the relatively low Native American ancestry in many of the samples, we lowered the inclusion threshold for this ancestral component to 3%. This allowed for the maximization of samples in the analysis while limiting the introduction of statistical noise resulting from individuals with very little Native American ancestry. For each continental ancestry, a reference panel of putatively ancestral sub-continental populations was built (see Table S1). The initial projection of Native American segments included 218 indigenous haplotypes derived from admixed samples, and 1,100 from the reference panel (i.e., all 493 individuals from [2], and 57 native Venezuelans). Aleutians, Greenlanders, native Venezuelans, and Surui were detected as the most extreme outliers in PCA space (see Figure S6), and were thus removed from subsequent ASPCA analyses. The final projection included a reference panel of 892 native haplotypes. The projection of European segments included 255 haplotypes from the admixed populations, and 2,882 from the reference panel (i.e., the subset of 1,387 European PopRes samples used in [3], plus 54 additional Iberian individuals sampled in Spain (Rodriguez et al., in revision). In this case we restricted to the same set of SNPs used in [3] (i.e., 192,821 sites after merging) in order to ensure the reproduction of the PCA map of Europe by Novembre et al. Finally, the projection of African segments included 55 haplotypes from the admixed populations, and 538 from the reference panel (i.e., 205 HapMap African samples: Yoruba from Nigeria and Kenyan Luhya; as well as 64 West African individuals from [4], including Kongo, Bamoun, Fang, and Igbo).

**Text S2. Measuring pairwise IBD between European and Latino populations**

To provide independent evidence on the sub-continental ancestry of European haplotypes derived from Caribbean Latinos, we considered segments that are identical by descent (IBD) between unrelated Latino individuals and a representative subset of European populations. We calculated a summary statistic, WELat, to inform about the proportion of shared DNA between pairs of populations. Specifically, WELat is the sum of lengths of all segments inferred to be shared IBD between a given European population "*E*" and Latino populations "*Lat*", normalized by sample size. When comparing WELat values for each Latino group across European populations, maximum pairwise IBD levels were observed in those pairs involving Spanish and, to a lesser extent, Portuguese samples (Figure S11), in agreement with our ASPCA results.